\documentclass[aps,prd,superscriptaddress,amsfonts,amssymb,amsmath,showpacs,nofootinbib,twocolumn, floatfix,longbibliography]{revtex4-1}

\usepackage[dvips]{graphicx}
\usepackage[squaren]{SIunits} 
\usepackage{aas_macros}
\usepackage{color}

\begin{document}

\title{Torsional oscillations of magnetized neutron stars with mixed poloidal-toroidal fields}

\author{Gibran H. de Souza} 
\email{e-mail: gibaifi@ifi.unicamp.br} 
\affiliation{Instituto de F\'isica Gleb Wataghin, UNICAMP, 13083-859, Campinas-SP, Brazil}

\author{Cecilia Chirenti} 
\email{e-mail: cecilia.chirenti@ufabc.edu.br} 
\affiliation{Centro de Matem\'atica, Computa\c c\~ao e Cogni\c c\~ao, UFABC, 09210-170 Santo Andr\'e-SP, Brazil}

\begin{abstract}
The quasiperiodic oscillations found in the three giant flares of soft gamma-ray repeaters observed to date have been interpreted as torsional  oscillations caused by a starquake related to a magnetospheric reconnection event. Motivated by these observations, we study the influence of the magnetic field geometry in the frequencies of the torsional oscillations of magnetized neutron stars. We use realistic tabulated equations of state for the core and crust of the stars and model their magnetic field as a dipole plus a toroidal component, using the relativistic Grad-Shafranov equation. The frequencies of the torsional modes are obtained by the numerical solution of the eigenvalue problem posed by the linear perturbation equations in the Cowling approximation. Our results show how the asteroseismology of these stars becomes complicated by the degeneracy in the frequencies due to the large relevant parameter space. However, we are able to propose approximately EOS-independent relations that parametrize the influence of the magnetic field in the torsional oscillations, as well as a testable scenario in which the rearrangement of the magnetic field causes an evolution of the frequencies. Finally, we show that there is a magnetic field configuration that maximizes the energy in the perturbation at linear order, which could be related to the trigger of the giant flare.  
\end{abstract}


\maketitle

\section{Introduction}
\label{sec:intro}

{\it Magnetars} are neutron stars powered by their usually very strong magnetic fields, that can reach  over $10^{15}$ G
\cite{2015RPPh...78k6901T}. Observations of soft gamma-ray repeaters (SGRs) seem to indicate that these objects are
magnetars, with high magnetic fields and low rotation rates \cite{1992ApJ...392L...9D,1995MNRAS.275..255T}. There have
been three giant flares associated with these objects observed so far, SGR 0526-66 in 1979
\cite{1979Natur.282..587M,1983A&A...126..400B}, SGR 1900+14 in 1998 \cite{1999Natur.397...41H} and SGR 1806-20 in 2004
\cite{2005Natur.434.1110T,2005Natur.434.1107P}. 

In the late-time tail of these events some quasiperiodic
oscillations (QPOs) were detected.
The QPOs observed in the giant flares of SGR 0526$-$66 (43.5 Hz) \cite{1983A&A...126..400B}, SGR 1900+14 (28, 54, 84
and 155 Hz) \cite{2005ApJ...632L.111S} and SGR 1806$-$20 (18, 26, 29, 92.5, 150, 625.5 and 1837 Hz)
\cite{2005ApJ...628L..53I,2006ApJ...637L.117W,2006ApJ...653..593S} (see also a study of some short recurring bursts performed by \cite{2014ApJ...795..114H}) seem to indicate that these could be characteristic
modes of oscillation of the stars that were excited by some catastrophic event, probably connected to the strong
magnetic fields of these neutron stars (see \cite{1996ApJ...473..322T} and \cite{2003MNRAS.346..540L} for two competing models that could explain the trigger of the giant flare).

However, the precise origin of these oscillations is still unclear 
(see \cite{2008ApJ...680.1398T} for a proposed description of the QPO emission), 
and many different descriptions of the oscillations have been proposed in
the literature. The initial simple model of crustal torsional oscillations \cite{1998ApJ...498L..45D} was met with
difficulties related to the coupling with Alfv\'en modes in the core \cite{2006MNRAS.368L..35L,2008MNRAS.385L...5S,2011MNRAS.410L..37G}, and more recently it
has been speculated that the geometry of the magnetic field could play an important role in the understanding of this
problem \cite{2016ApJ...823L...1L}. 

Further complicating the problem, details of the crust can also change the analysis, see for instance \cite{2014PhRvC..90b5802D}. Other works have also included effects such as a superfluid or a pasta phase in the stellar core, see \cite{2018MNRAS.476.4199G,2018MNRAS.479.4735S} and other references therein.

Moreover, even the observational situation is not clear. Although claims have been made for QPOs at a variety of frequencies, the strength, significance and duration of each particular QPO remain uncertain. As an example, the 625 Hz QPO of SGR 1806-20 was studied in detail in \cite{2014ApJ...793..129H}, where the analysis of the corresponding RXTE data concluded that the QPO was probably present in the signal for only 0.5 s.  Additionally, a new analysis of the giant flares done by \cite{2018A&A...610A..61P} did not confirm te previously reported QPOs, but found new frequencies. More recently, a reanalysis of the data performed by \cite{Miller:2018kmk} presented evidence for \emph{no} long lived QPOs, favoring instead a scenario with several re-excitations of the oscillations. 

In this paper we study how the addition of a toroidal component to the dipolar magnetic field can modify the frequencies of the torsional modes of the crust. 
We do not aim here to present a model that can reproduce all of the observed frequencies for each star, for this would require some careful fine tuning of the stellar properties.
Rather, we work with a comprehensive set of core and crust equations of state, and explore the parameter space in compactness, magnitude and geometry of the magnetic field. 

This paper is organized as follows. In Section \ref{sec:models} we present our equilibrium models for the magnetized neutron stars. In Section \ref{sec:results} we discuss our analytical and numerical setup for solving the perturbation equations and obtaining the quasinormal frequencies in the relativistic Cowling approximation and we find some nearly EOS-independent relations followed by the frequencies. In Section \ref{sec:evolution} we consider a quasi-static evolution scenario for the magnetic field of the neutron star and its consequences as a possible trigger for the giant flare. We present our final remarks in Section \ref{sec:conclusions}.

\section{Equilibrium models with mixed magnetic field configurations}
\label{sec:models}

In order to describe a magnetized neutron star, we use the general stress energy tensor given by 
\begin{equation} 
T^{\alpha\beta}=(p+\rho+H^{2})u^{\alpha}u^{\beta}+\left(p+\frac{H^{2}}{2}\right)g^{\alpha\beta}-H^{\alpha}H^{\beta},
\label{eq:stress-energy}
\end{equation}
where $p$ is the fluid pressure, $\rho$ is the total energy density, $u^{\alpha}$ is the fluid 4-velocity, $g^{\alpha\beta}$ is the spacetime metric and
$H^{\alpha} = B^{\alpha}/4\pi$ is the magnetic field. It is a well-known fact that a very strong magnetic field can deform the spherical symmetry of the star. Nevertheless, even at magnetar-type field strengths of order $10^{15}$ G the deformation is negligible, as the magnetic energy is much less than 1\% of the gravitational energy of the star \cite{2007MNRAS.375..261S}. Therefore we can add the magnetic field as a linear perturbation on the background spacetime of our slowly rotating magnetized neutron star described by a line element of the form
\begin{equation}
ds^{2}=-e^{\nu}dt^{2}+e^{\lambda}dr^{2}+r^{2}d\theta^{2}+r^{2}\sin^{2}\theta \left[d\phi-\omega dt\right]^{2},
\label{eq:metric}
\end{equation}
where $\nu$, $\lambda$ and $\omega$ are functions only of the radial coordinate $r$. However, SGRs are very slow rotators and therefore we can set $\omega = 0$ (as the frame dragging is negligible for these stars) recovering a spherically symmetric line element.

We describe the hydrostatic equilibrium and internal structure of our relativistic magnetized neutron stars with the Tolman-Oppenheimer-Volkoff (TOV) equations \cite{1939PhRv...55..364T,1939PhRv...55..374O}. We use three different equations of state (EOSs) for the core and three different EOSs for the crust (each crust has also a different core-crust transition density) as summarized in Tables \ref{tableEOScore} and \ref{tableEOScrust}. 

\begin{figure}
\begin{center}
\includegraphics[width=0.95\linewidth]{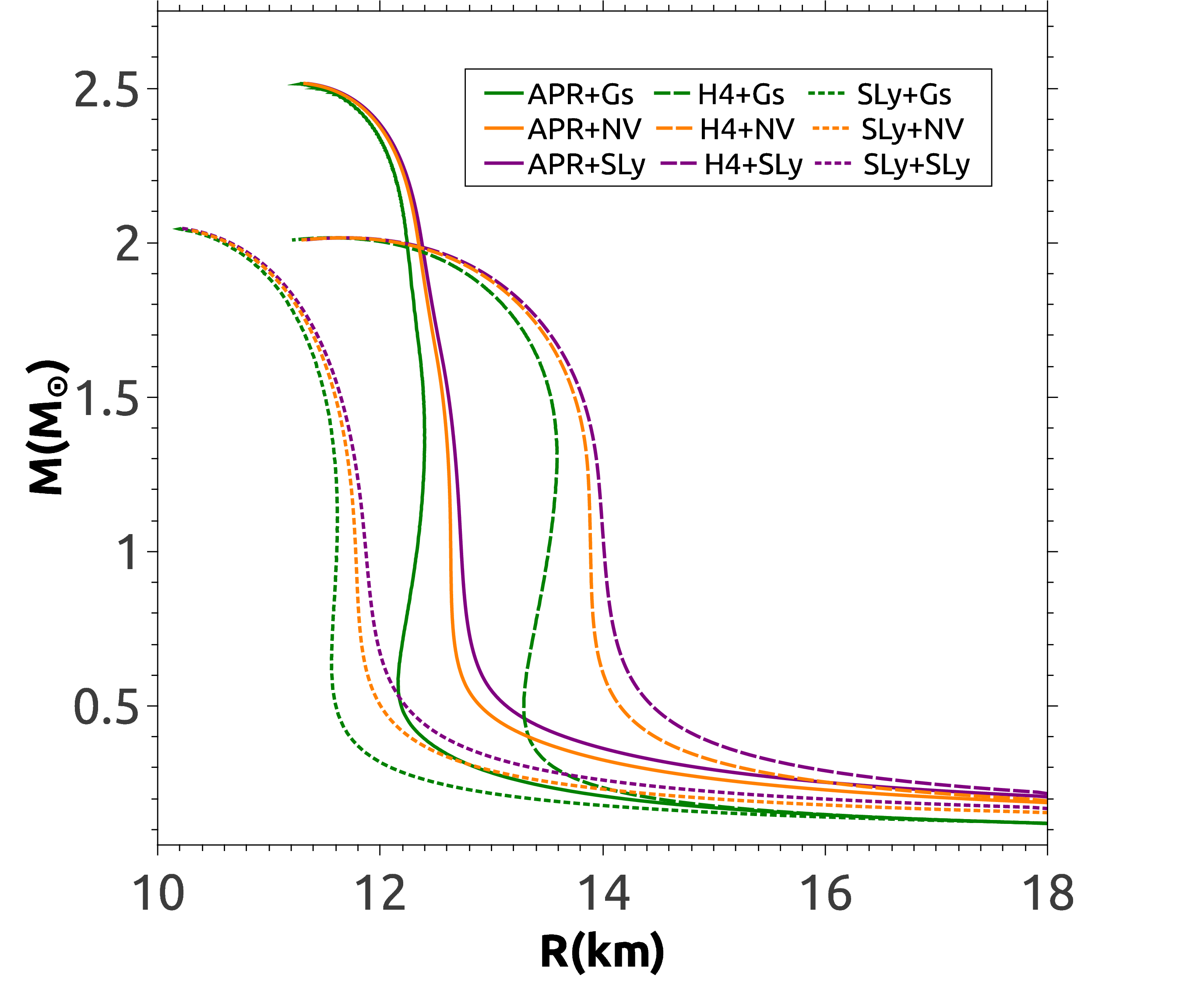}
\end{center}
\caption{Mass-radius relation for the crust+core equations of state used in our study. The core EOS determines the bulk properties of the star, but the crust EOS will be the most important ingredient for the characteristics of the torsional oscillations.}
\label{fig:M-R}
\end{figure}

The mass-radius $M-R$ relation for the resulting equilibrium configurations is presented in Fig. \ref{fig:M-R}. As we can see in the figure, the total mass and radius of the star are mostly determined by the core EOS. The crust EOS results in a correction on the radius of less massive configurations. For a 1.4 $M_{\odot}$ star, our choices for the crust EOS imply a variation of $\approx 1-2\%$  on the total radius of the star. However, the choice of the crust EOS will be the most important ingredient to determine the properties of the torsional oscillations of the star, as we will show in Section \ref{sec:results}. 

\begin{table}
\caption{The neutron star core equations of state used in this work.}
\label{tableEOScore}
  \begin{tabular}{lcllc}
  \hline
     core EOS & Ref. \\     
     \hline
     APR  & \cite{1998PhRvC..58.1804A} \\
     H4    & \cite{2006PhRvD..73b4021L} \\
     SLy   & \cite{2001AA...380..151D} \\
     \hline
 \end{tabular}
\end{table}

\begin{table}
\caption{The neutron star crust equations of state used in this work, and the core-crust transition density $\rho_c$ used for each crust.}
\label{tableEOScrust}
  \begin{tabular}{lcllc}
  \hline
     crust EOS &  $\rho_c$ [g/cm$^3]$ &Ref.\\
     \hline
     Gs   & $2.01 \times 10^{14}$ & \cite{2012PhRvC..85e5804S} \\
     NV  & $2.40 \times 10^{14}$   & \cite{1973NuPhA.207..298N}\\
     SLy & $1.34 \times 10^{14}$  & \cite{2012PhRvC..85e5804S} \\
     \hline
 \end{tabular}
\end{table}

\begin{figure}
\begin{center}
\includegraphics[width=0.95\linewidth]{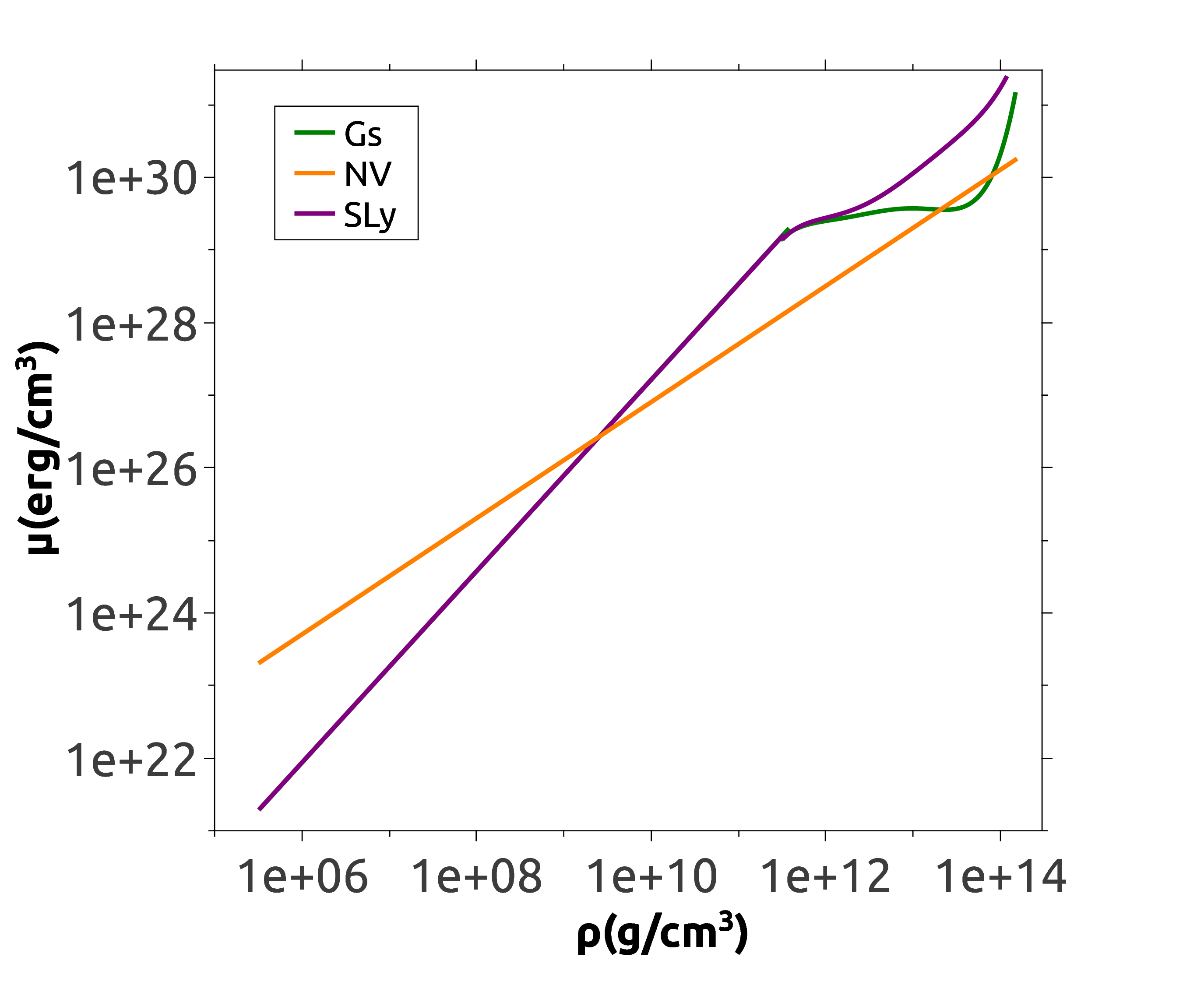}
\end{center}
\caption{Polynomial fits for the shear modulus $\mu$ as a function of the density $\rho$ for our set of crust EOSs.}
\label{fig:shear}
\end{figure}

The shear modulus $\mu$ of the crusts we are considering will also be needed in our analysis in Section \ref{sec:results}. We follow \cite{2007MNRAS.375..261S} and calculate the shear modulus in the zero  temperature limit of eq. (13) from \cite{1991ApJ...375..679S}. Our results for the shear modulus as a function of the density $\rho$ are given in Fig. \ref{fig:shear}, where we present  smooth composite polynomial fits for crusts Gs and SLy, which also describe the change of behavior observed for $\rho \approx 3.8 \times 10^{11}$ g/cm$^3$, and for the older crust model NV we use a polynomial fit provided by \cite{1998ApJ...498L..45D}.

When considering possible magnetic field configurations for our models, we must remember that it has long been known that stars with purely toroidal magnetic field configurations are always unstable due to Parker and/or 
Tayler instabilities \cite{1973MNRAS.161..365T} (see also \cite{2008PhRvD..78b4029K,2011A&A...532A..30K}), and that the purely poloidal field case is also unstable due to ``sausage'' or ``kink" instabilities 
(se for example \cite{1973MNRAS.163...77M,1974MNRAS.168..505M,1977ApJ...215..302F,2011ApJ...735L..20L,2012ApJ...760....1C}). A stable configuration requires a mixed
poloidal-toroidal field inside the star, and the appearance of the field at the surface is a dipole with smaller
contributions from higher multipoles \cite{2006A&A...450.1077B,2008MNRAS.386.1947B}. 

In order to describe a relativistic neutron star with a general magnetic field configuration, including both poloidal
and toroidal components, we use the TOV equations to
provide the metric coefficients of the spacetime and the fluid variables inside the star, and the relativistic Grad-Shafranov
equation \cite{Ioka:2003nh} to describe the magnetic field. The Grad-Shafranov equation is derived from the Maxwell equations solved in the stellar background described by the TOV equations. The vector potential is given by \cite{2008MNRAS.385.2080C}
\begin{equation}
\label{eq:a}
e^{-\lambda}a_1'' + \frac{\nu' - \lambda'}{2}e^{-\lambda}a_1' + \left(\zeta^2 e^{-\nu} - \frac{2}{r^2}\right)a_1 = 
4\pi(\epsilon+p)r^2c_0\,,
\end{equation}
where the constant $\zeta$ is the ratio between the poloidal and toroidal magnetic
field components, $a_1(r)$ is the radial profile of the vector potential and $c_0$ is a constant that gives the amplitude of the source term (current) on the r.h.s. of eq. (\ref{eq:a}). This single equation describes a mixed poloidal-toroidal magnetic field configuration because we have  assumed a current given by two separate contributions that act as the source of the poloidal and toroidal magnetic field components; they are, respectively, proportional to the source term on the r.h.s. of eq. (\ref{eq:a}) and to the $\zeta^2$ term on the l.h.s. of the same equation.

In order to obtain $a_1(r)$ from the center of the star at $r=0$ to the exterior spacetime we follow the standard procedure and match the regular series expansion of the solution near the center to the numerical solution of eq. (\ref{eq:a}) up to the surface at $r = R$, where we require that  both $a_1(r)$ and its first derivative match continuously the vacuum solution outside the star, where all sources are zero (in particular, $\zeta = 0$ outside the star) and eq. (\ref{eq:a}) has an analytical solution. A more detailed description of this method is given for instance by \cite{2013PhRvD..88j4018C}. It is however worth noting that this is a simplification of the problem and the solution of the Grad-Shafranov equation becomes much more involved when a description of the magnetosphere is included, see for example the works of \cite{2015MNRAS.447.2821P} and \cite{2018MNRAS.474..625A}.

In this formalism we also assume that the magnetic field is not strong enough to deform the
spherical symmetry of the star \cite{2008MNRAS.385.2080C,2009MNRAS.397..913C} and the magnetic field components are given by: 
\begin{subequations}
\begin{align}
\label{eq:B1}
B_r &= \frac{2e^{\frac{\lambda}{2}}a_1 \cos\theta}{r^2}\,,\\
B_{\theta} &= -e^{-\frac{\lambda}{2}}a_1' \sin \theta\,,\\
B_{\phi} &= -\zeta e^{-\frac{\nu}{2}}a_1\sin\theta\,.
\label{eq:B3}
\end{align}
\end{subequations}
Here we stress that, although the poloidal components $B_r$ and $B_{\theta}$ extend to infinity, the toroidal component is zero outside the star, where we set $\zeta = 0$. Strictly speaking, this causes a discontinuity in $B_{\phi}$ and the appearance of a surface current. However, this discontinuous behavior is only a consequence of our simplified treatment. In a real magnetized neutron star, this discontinuous jump would be continuously spread out in the magnetosphere to match the vacuum solution. Moreover, it does not affect our calculations, which are restricted to the interior of the star.

\begin{figure*}
\begin{center}
\includegraphics[angle=-90,width=0.8\linewidth]{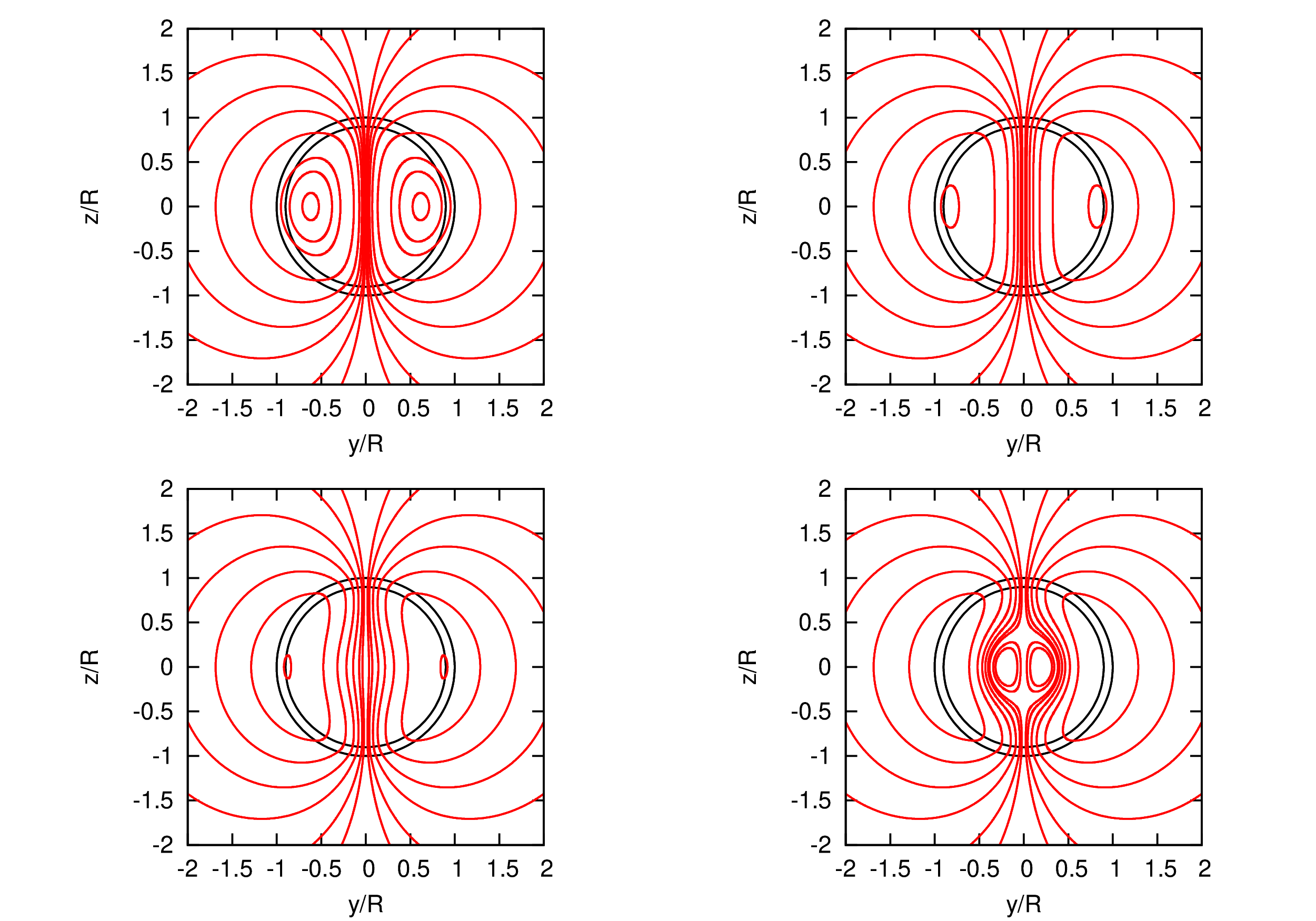}
\end{center}
\caption{Magnetic field lines for typical configurations with a dipolar field and an increasing toroidal field component. All four stars have the APR+Gs EOS, $M = 1.4 M_{\odot}$ and magnetic field amplitude $B = 10^{15}$ G at the pole. The outer black circle indicates the surface of the star, while the inner circle indicates the base of the crust. The ratio $\zeta$ between the poloidal and toroidal components is 0.00 (top left), 0.26 (top right), 0.28 (bottom left) and 0.30 (bottom right). We restrict our range for the parameter $\zeta$ for each EOS in order to prevent the formation of disjoint domains for the magnetic field lines as shown in the bottom right plot. 
}
\label{fig:Bfield}
\end{figure*}

\begin{figure*}
\begin{center}
\includegraphics[angle= 0,width=0.4\linewidth]{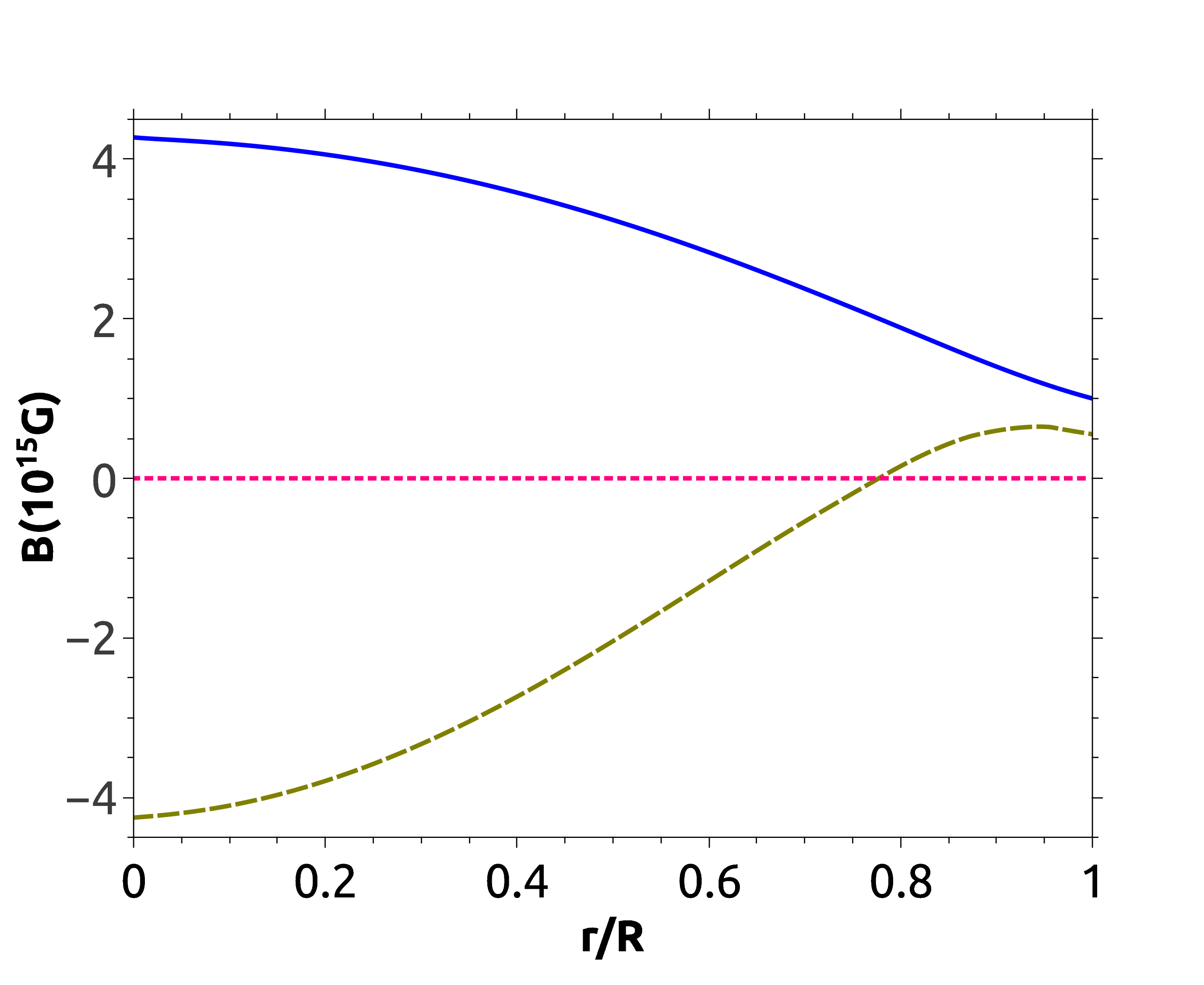}
\includegraphics[angle= 0,width=0.4\linewidth]{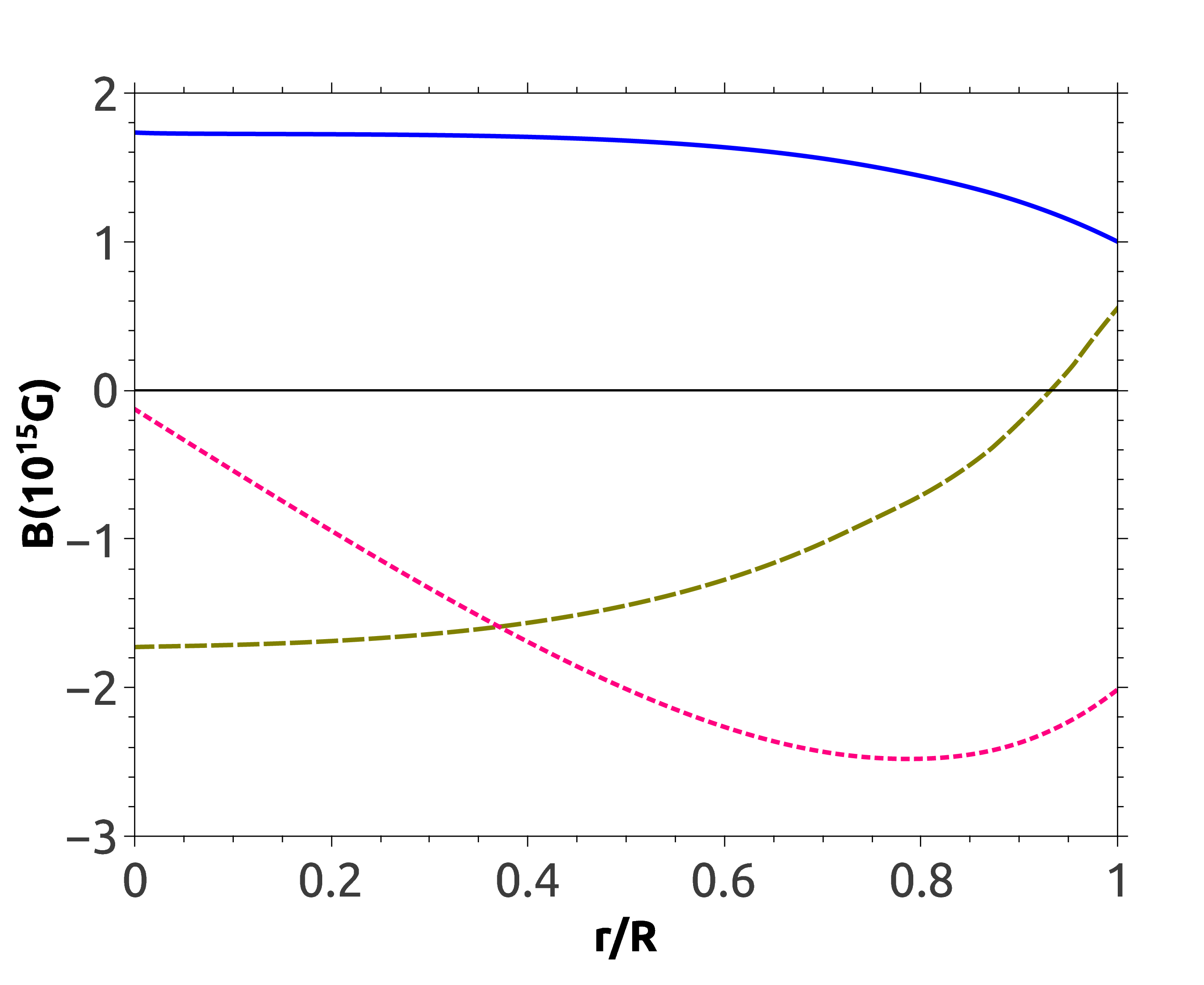}
\includegraphics[angle= 0,width=0.4\linewidth]{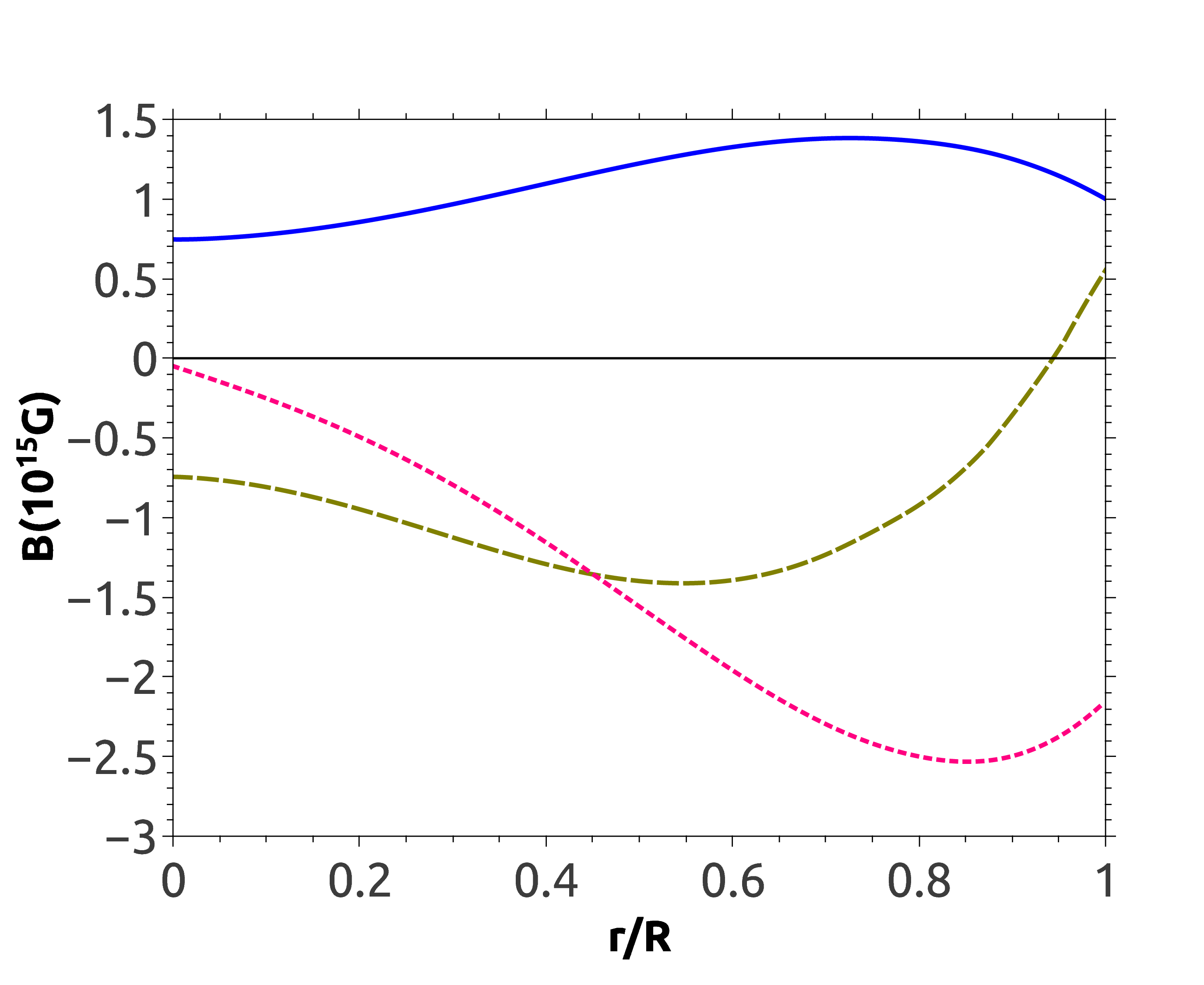}
\includegraphics[angle= 0,width=0.4\linewidth]{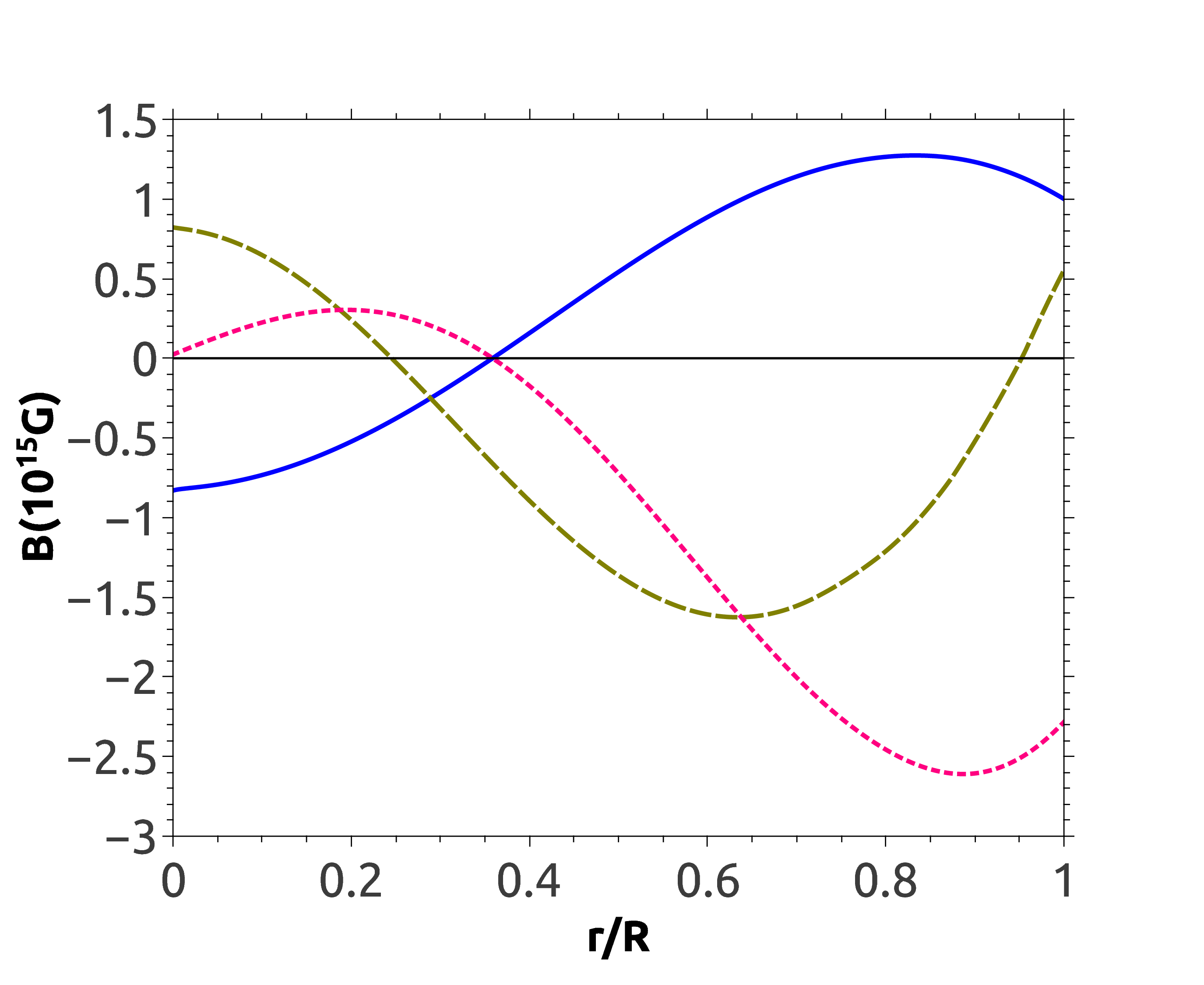}
\end{center}
\caption{Radial profiles of the magnetic field components $B_r$ (at the axis $\theta = 0$), $B_{\theta}$ and $B_{\phi}$ (at the equatorial plane $\theta  =\pi/2$), represented with solid, dashed and dotted lines, respectively, calculated inside the star with eqs. (\ref{eq:B1})-(\ref{eq:B3}) for the same configurations presented in Fig. \ref{fig:Bfield}. The poloidal components $B_r, B_{\theta}$ extend continuously outside the star, but the toroidal component $B_{\phi}$ is set to zero at the surface. The configuration with higher $\zeta$ and disjoint domains for the field lines presents a node in the $B_r$ component inside the star, as can be seen in the bottom right panel.
}
\label{fig:Br}
\end{figure*}

We work with a sequence of static magnetic field configurations, starting with a purely dipolar field and then supplementing it with an increasing toroidal component $B_{\phi}$, obtained by increasing the parameter $\zeta$. Typical examples of the magnetic field lines are presented in Fig. \ref{fig:Bfield}. As we can see in the sequence of panels with increasing toroidal field in the figure, the purely dipolar field lines are distorted by the presence of the toroidal field to the point where the field lines are defined in disjoint domains in the bottom right panel \cite{2008MNRAS.385.2080C}. This can also be seen in Fig. \ref{fig:Br}, where we present the radial profiles of the magnetic field components $B_r, B_{\theta}$ and $B_{\phi}$ (with solid, dashed and dotted lines, respectively) given by eqs. (\ref{eq:B1})-(\ref{eq:B3}) for the configurations presented in Fig. \ref{fig:Bfield}. Here we can see (again in the bottom right panel) that the configuration with disjoint domains for the field lines has a node in the $B_r$ component inside the star.

For each crust+core EOS used in our work, we determine the value of $\zeta_{\rm{max}}$ where this node appears and restrict our range to $\zeta < \zeta_{\rm{max}}$. The influence of the toroidal field component on the torsional oscillations will come indirectly through this deformation of the poloidal field components, as we will show in our discussion of the linear perturbation equations in Section \ref{sec:results} below. It is also important to note that our linear analysis will take the magnetized stellar equilibrium configuration as a fixed background on which the perturbation equations are solved. Therefore, the nonlinear stability analysis of the magnetic field (see for example the analysis of \cite{2011ApJ...735L..20L,2012ApJ...760....1C}) is out of the scope of our work.

\section{Influence of the toroidal magnetic field component in the torsional oscillations of the crust}
\label{sec:results}

Working in the Cowling approximation \cite{1941MNRAS.101..367C} (in which we neglect all metric perturbations) and considering only barotropic
perturbations (for a review on neutron star quasinormal modes, see \cite{1999LRR.....2....2K}), we have 10 perturbed variables for a non-rotating neutron star: ~$\delta \rho, \delta p, \delta
u^{\alpha}, \delta H^{\alpha}$.
The linear perturbation equations can be obtained by manipulations of the perturbed Euler equation (the momentum conservation equation), which is obtained from the projection of the conservation equation of the stress energy tensor in the direction orthogonal to the fluid 4-velocity (with $i=r,\theta,\phi$),
\begin{equation}
\delta \left( \{\delta^{i}_{\alpha}+u^{i}u_{\alpha}\}T^{\alpha\beta}_{~;\beta}\right)=0,
\end{equation}
and the perturbed energy conservation equation (given by the projection parallel to the fluid 4-velocity),
\begin{equation}
\delta\left(u_{\alpha}T^{\alpha\beta}_{~;\beta}\right)=0,
\end{equation}
together with the perturbed induction equation, which is derived from the homogeneous Maxwell equations $F_{[\alpha \beta; \gamma]} = 0$ (where $F_{\alpha\beta}$ is the Faraday tensor) and can be written as
\begin{equation}
\label{eq:induction}
\delta\left\{(u^{\alpha}H^{\beta}-H^{\alpha}u^{\beta})_{;\beta}\right\}=0.
\end{equation}
Using the EOS, which gives $p = p(\rho)$, together with constraints given by the ideal MHD approximation and the 4-velocity normalization condition, we can reduce the number of independent variables to 7: ~$\delta p, \delta u^{i}, \delta H^{i}$. The formalism used for deriving the perturbation equations is known in the literature and we refer the reader to the derivation presented by \cite{2001MNRAS.328.1161M}. This formalism was  used in several other works, including the calculation of r-modes of magnetized neutron stars in \cite{2013PhRvD..88j4018C}, where a detailed derivation of the perturbation equations is given.

Here we restrict our study to the torsional modes of oscillation of the crust of a magnetized neutron star, and we analyze the influence of the magnetic field geometry on the mode frequencies. Therefore we neglect the coupling of the crustal modes to the fluid modes in the core \cite{2006MNRAS.368L..35L} and the coupling between axial and polar modes \cite{2012MNRAS.423..811C}. It is important to remark here that a full treatment of the coupled oscillations would require the study of global magneto-ellastic oscillations \cite{2009MNRAS.396.1441C,2011MNRAS.410.1036V,2011MNRAS.410L..37G,2011MNRAS.414.3014C}. 

Our choice of neglecting the coupling to the core and studying only the spectrum of the crust is motivated by the recent results of \cite{Miller:2018kmk}. Their analysis presented a picture consistent with short lived QPOs, which is in agreement with the scenario originally proposed by \cite{2006MNRAS.368L..35L} of crustal modes being quickly damped by their coupling to a continuum of modes in the core. This quick damping has been viewed as a strong objection against the simple model of crustal oscillations for explaining the QPO frequencies because the QPOs were thought to be long lived, in which case the full problem has to be studied and the gaps in the continuum are driving the oscillations. This is no longer a difficulty if the QPOs do not last for more than 1 s, as suggested by \cite{Miller:2018kmk}. Moreover, although the \emph{qualitative} picture is described by the coupling to the modes in the core, the \emph{quantitative} calculation of the frequencies can be done neglecting it, since the damping time is long compared to the period of typical torsional oscillations with frequencies of at least approximately 20 Hz. Therefore the observed frequencies could, in principle, be explained \emph{by the crustal modes alone}. In light of these results it is valuable to understand how the crustal modes depend on the details of the star, including the EOS, compactness, magnetic field strength and geometry.

Our treatment results in a considerable simplification of the problem, but interesting results for the torsional crustal frequencies can be achieved, and the trends we observe are expected to be independent of the neglected couplings to other modes. Although the toroidal component of the magnetic field $B_{\phi}$ does not appear explicitly in the linear perturbation equations in our treatment, it still influences the results by inducing changes in the poloidal components $B_r$  and $B_{\theta}$ of the magnetic field through the $\zeta$ parameter in the radial function of the vector potential $a_1$, see eqs. (\ref{eq:a}) and (\ref{eq:B1})-(\ref{eq:B3}). This procedure is similar to the one carried out by \cite{2008MNRAS.385.2161S}; although they were mostly concerned with a very different magnetic field configuration (all field components expelled from the core and confined in the crust by the presence of a type Ia superconductor core), we were able to compare our results with an exploratory analysis for configurations more similar to ours presented in their appendix, with very good agreement (see below in this section).

The only perturbed fluid
variable present in the torsional modes of oscillation is, to first order, $\delta u^{\phi}$, the perturbation in the azimuthal component of the 4-velocity of the stellar fluid \cite{1983MNRAS.203..457S}, to
which we must add the $\delta H^i$ perturbations in the magnetic field components, in the case of magnetars \cite{1998ApJ...498L..45D}. \footnote{It is important to remark that the linearized perturbations $\delta H^i$ of the stellar magnetic field are first order in the equilibrium magnetic field and first order in the amplitude of the stellar fluid oscillation.}

In a nutshell, both the non-vanishing components of the linearized stress shear tensor \cite{1983MNRAS.203..457S} and the $\delta H^i$ components of the perturbed magnetic field (given by the linearized induction equation (\ref{eq:induction})) can be described in terms of $\delta u^{\phi}$, which can be written as
\begin{equation}
\delta u^{\phi} = e^{-\frac{\nu}{2}}{\cal Y}_{,t}\frac{(P_{l}(\cos \theta))_{,\theta}}{\sin \theta}\,,
\label{eq:delta_u_phi}
\end{equation}
where ${\cal Y}(t,r) = e^{i\omega t}{\cal Y}(r)$ and $P_l(\cos \theta)$ is the Legendre polynomial of order $l$. The final radial perturbation equation for ${\cal Y}(r)$ can be put in the form
\begin{equation}
{\cal A}_l(r){\cal Y}_{,rr} + {\cal B}_l(r){\cal Y}_{,r}  + {\cal C}_l(r){\cal Y} = 0\,,
\label{eq:Y}
\end{equation}
where the coefficients are given in terms of the functions describing the equilibrium metric, fluid and magnetic field of the star as
\begin{widetext}
\begin{align}
\label{eq:coeffA}
{\cal A}_l(r) &= \mu+ \frac{(1+2\lambda_1)(a_1)^2}{\pi r^4}\,,\\
\label{eq:coeffB}
{\cal B}_l(r) &= \left(\frac{4}{r} +\frac{\nu' - \lambda'}{2}\right)\mu +\mu' + 
 \frac{(1+2\lambda_1)a_1}{\pi r^4}\left[ \left(\frac{\nu' - \lambda'}{2}\right)a_1 + 2a_{1,r}\right]\,,\\
{\cal C}_l(r) &= 
\left[\left(\epsilon + p + \frac{(1+2\lambda_1)(a_1)^2}{\pi r^4}\right)e^{\lambda} - \frac{\lambda_1(a_{1,r})^2}{2\pi r^2}\right]\omega^2e^{-\nu} 
- (\lambda - 2)\left( \frac{\mu e^{\lambda}}{r^2} - \frac{\lambda_1(a_{1,r})^2}{2\pi r^4}\right) + 
\nonumber\\
&+ \frac{(2+5\lambda_1)a_1}{2\pi r^4}\left[ \left(\frac{\nu' - \lambda'}{2}\right)a_{1,r} + a_{1,rr}\right]\,,
\label{eq:coeffC}
\end{align}
\end{widetext}
with $\lambda = l(l+1)$, $\lambda_1 = -l(l+1)/((2l-1)(2l+3))$ and we have neglected the $l\pm2$ couplings that appear when we use eq.(\ref{eq:delta_u_phi}) to attempt to separate variables \cite{2007MNRAS.375..261S}. This simplification is physically expected to provide reliable results for the eigenfrequencies as long as the corrections caused by the magnetic field are small compared with the frequencies, which is true for our main results as we will see below. Moreover, this procedure was explicitly shown to give very accurate results when compared with 2D numerical evolutions of the complete perturbation equations for magnetic field strengths of up to $8 \times 10^{14}$ G (see Fig. 3 of \cite{2008MNRAS.385.2161S}). We solve this equation with a shooting method, imposing the zero traction condition at the base of the crust, $r = R_{\rm c}$, and zero torque at the surface of the star, $r = R$.

\begin{figure}
\begin{center}
\includegraphics[width=0.95\linewidth]{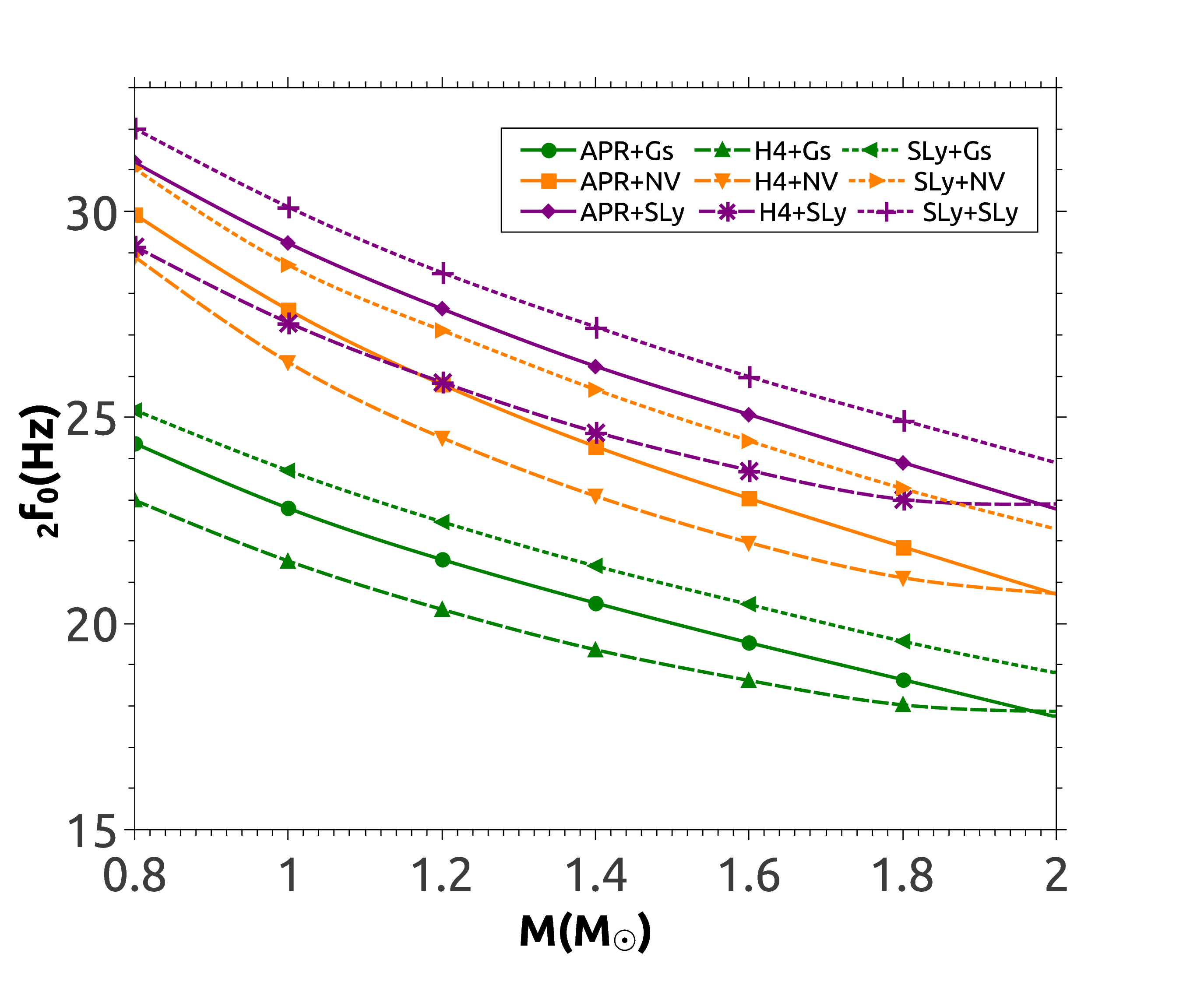}
\includegraphics[width=0.95\linewidth]{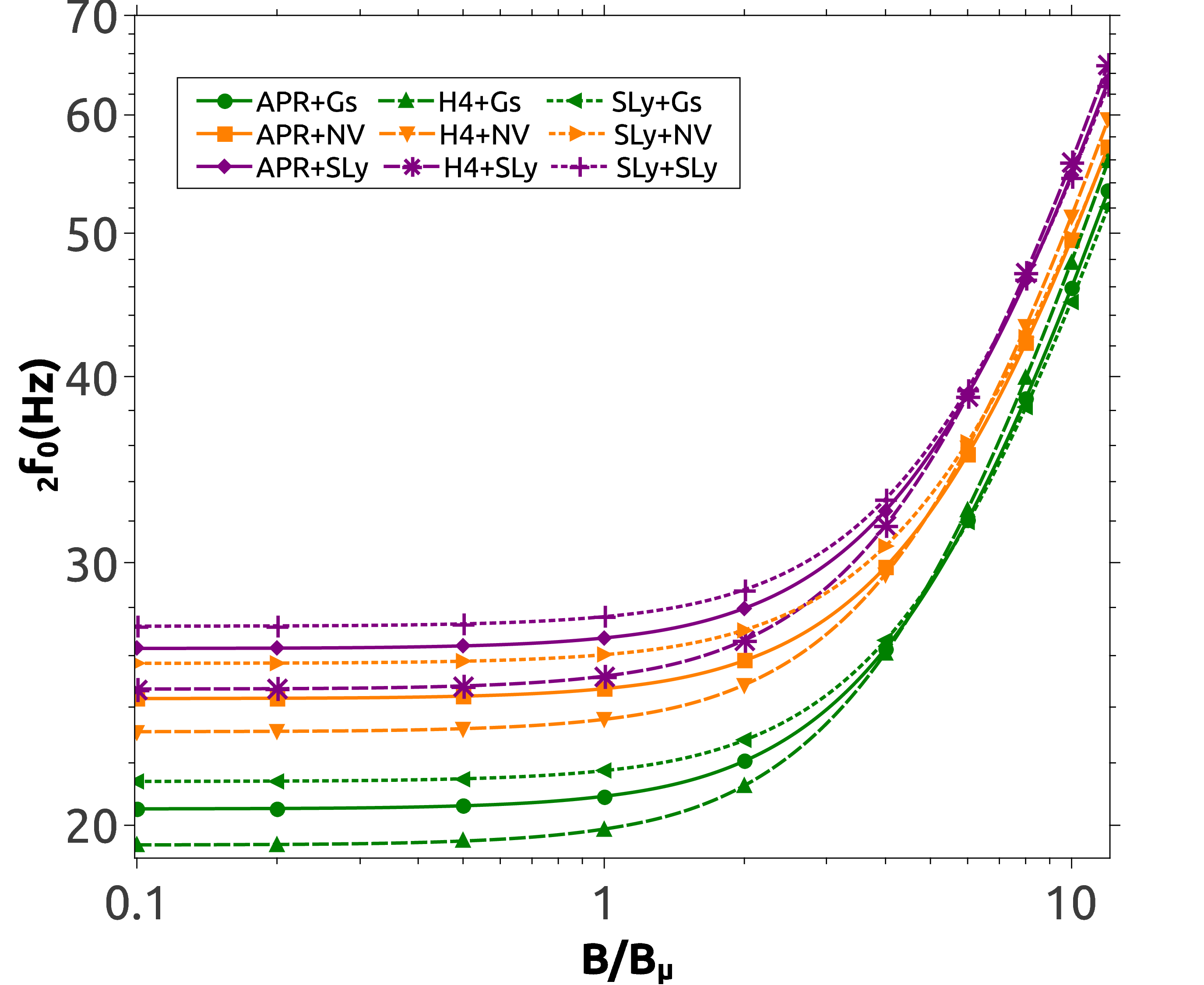}
\end{center}
\caption{Upper plot: Frequency $_2f_0$ of the fundamental ($n=0$) $l = 2$ mode of torsional oscillations for non-magnetized stars with different masses. 
Lower plot: Same as the upper plot, but for magnetized stars with a fixed mass $M = 1.4 M_{\odot}$ as a function of the magnetic field strength at the pole $B$, normalized by $B_{\mu} = 4 \times 10^{15}$ G. }
\label{fig:tests}
\end{figure}

Initially we tested our code for non-magnetized stars, generating the results presented in the upper plot of Fig. \ref{fig:tests}. Our results were compared with those of \cite{2007MNRAS.375..261S} for the frequency $_lf_n$ of the fundamental ($n=0$) mode with $l = 2,3,\ldots, 10$ for their A core EOS with an SLy crust and different masses. We found that in all cases our results agreed within less than 3\%, which can be explained by the slightly different value of density we used for the core-crust transition density $\rho_c$ (see Table \ref{tableEOScrust}), but also different samplings or interpolations of the tabulated EOS can produce some small differences. 
Figure \ref{fig:tests} shows that the SLy crust always presents the highest values for the frequencies, followed by the NV and the Gs crusts; but for a given crust EOS, the SLy core presents the highest frequencies, followed by the APR and H4 cores. The difference in the frequencies can be quite large: for a $1.4 M_{\odot}$ star, the difference in the frequencies across the different core+crust EOS combinations reaches $\approx 45\%$.
We note that the mode frequency, even in the simple non-magnetized case, is already degenerate with the mass and composition of the star. 

For our second test we included a purely dipolar magnetic field, see the results in the lower plot of Fig. \ref{fig:tests}. We have confirmed results from \cite{2007MNRAS.375..261S}, which proposed after \cite{1998ApJ...498L..45D} that the frequencies should depend on the magnetic field strength $B$ as
\begin{equation}
_lf_n = \phantom{ }_lf_n^0\left[1 + \phantom{ }_2\alpha_0\left(\frac{B}{B_{\mu}}\right)^2\right]^{1/2}\,,
\label{eq:f}
\end{equation}
where $_lf_n^0$ is the value of the frequency for a non-magnetized star, $B_{\mu} = 4 \times 10^{15}$ G and the coefficient $\phantom{ }_2\alpha_0$ is of order unity and depends on the details of the star. We compared our result for $\phantom{ }_2\alpha_0$ with the $1.4 M_{\odot}$ A+SLy star from \cite{2007MNRAS.375..261S} and found that our results agree within 5.5\%.

\begin{figure*}
\begin{center}
\includegraphics[width=0.3\linewidth]{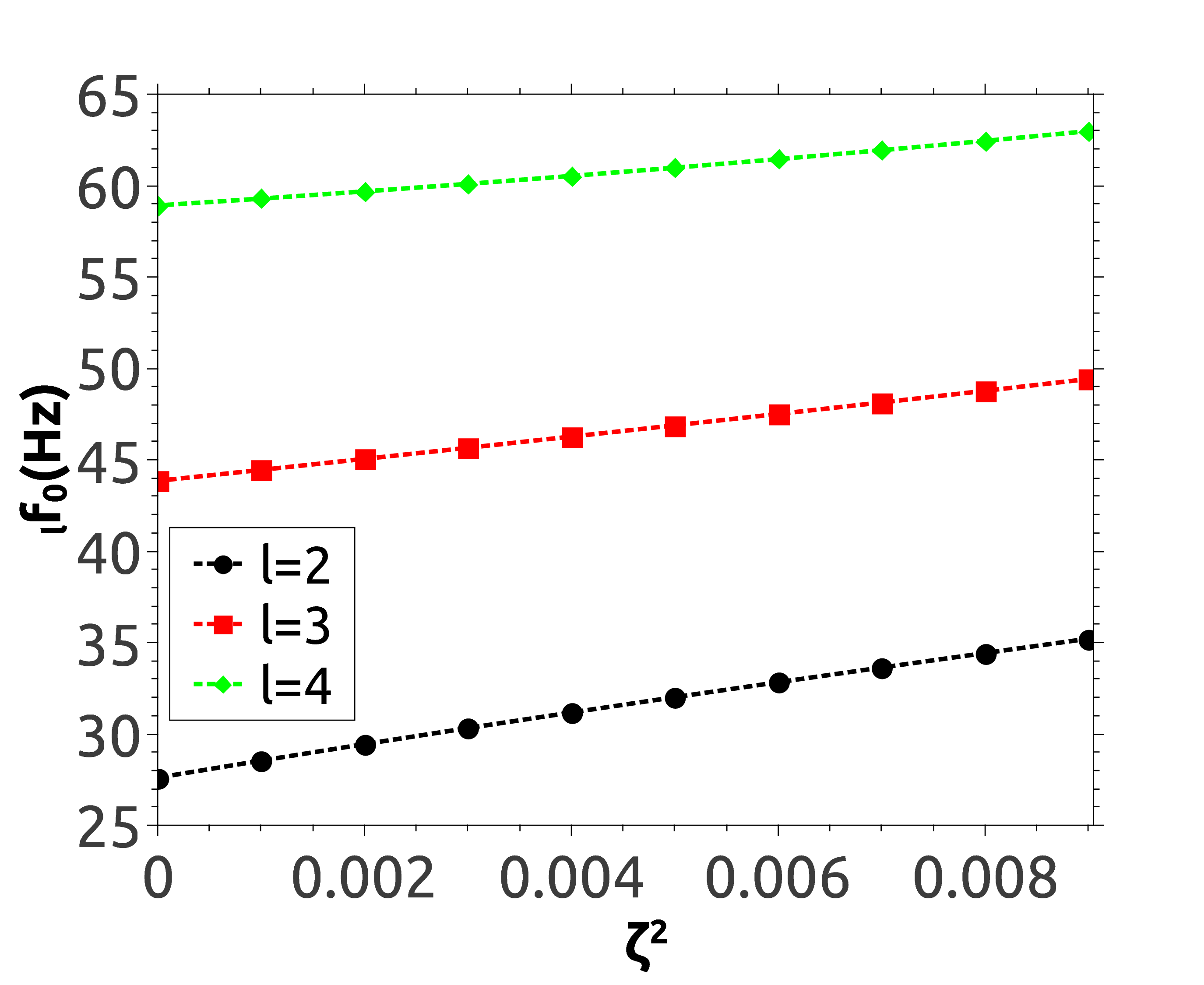}
\includegraphics[width=0.3\linewidth]{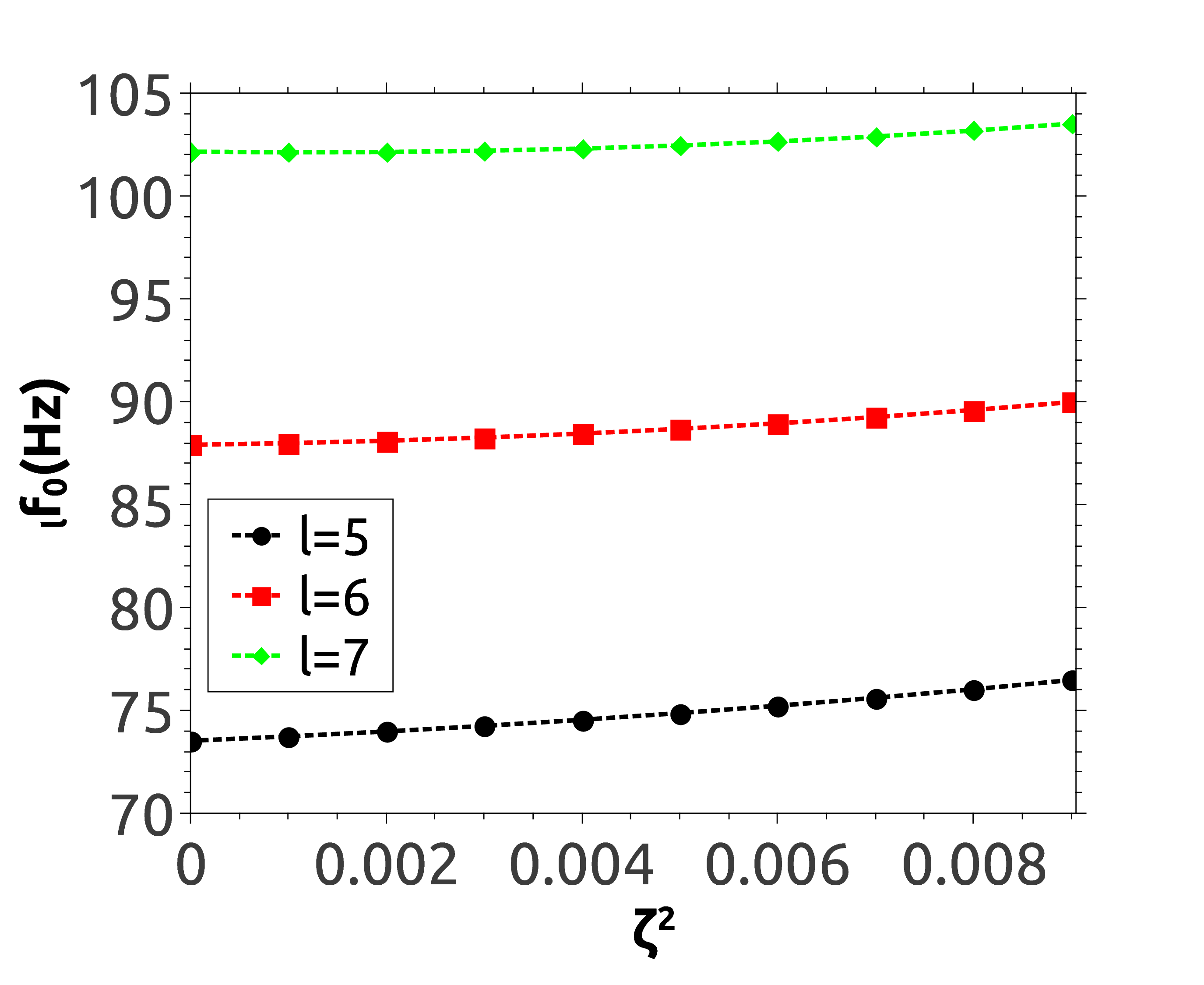}
\includegraphics[width=0.3\linewidth]{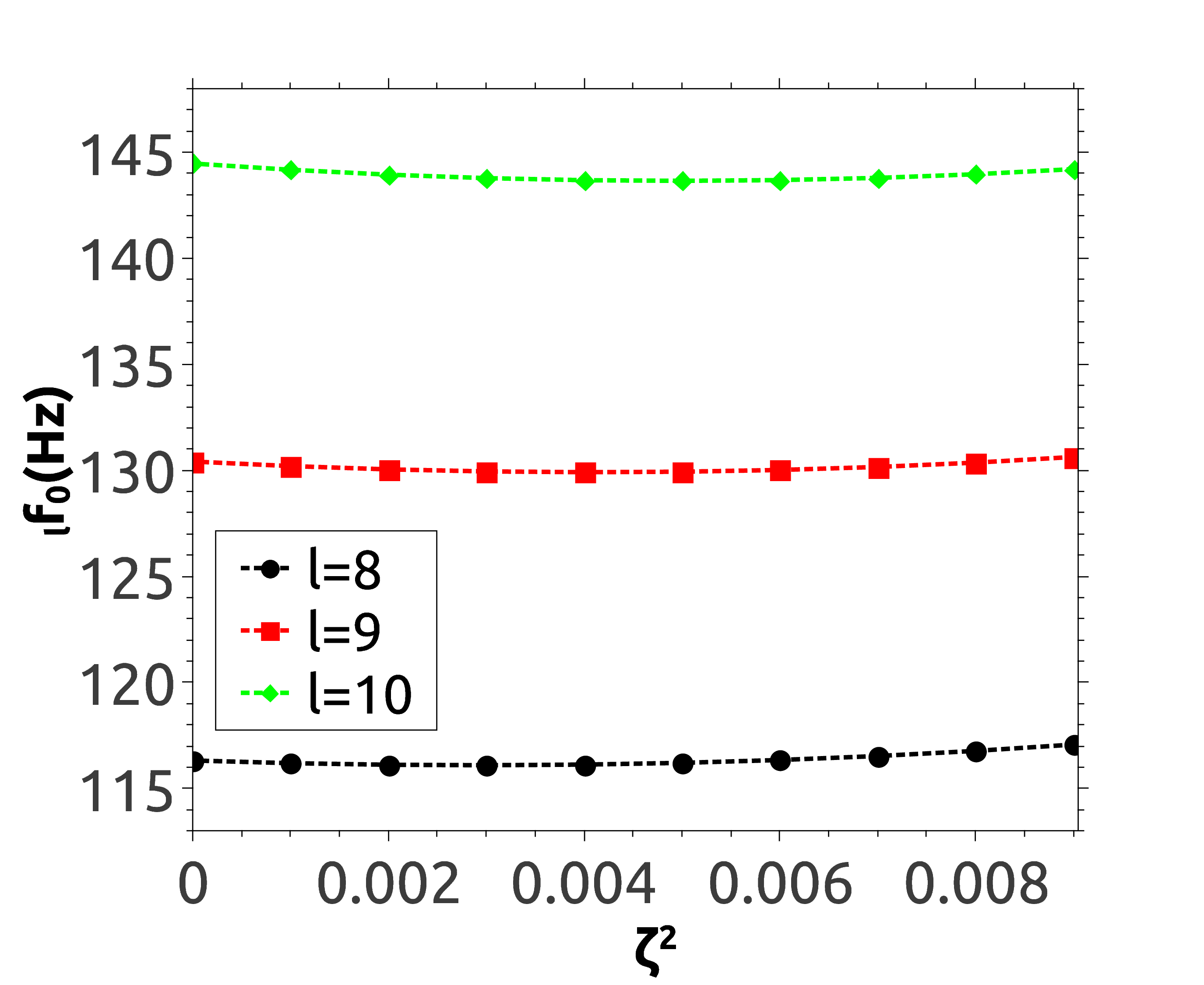}
\end{center}
\caption{Influence of the magnetic field geometry on the frequencies of the first modes of torsional oscillations for a sequence of 1.4 $M_{\odot}$ SLy+SLy stars. The sequence has a fixed amplitude $10^{15}$ G of the magnetic field at the pole and increasing toroidal field starting with a pure dipole configuration at $\zeta = 0$. The lowest modes are more sensitive to $\zeta$, but the effect of increasing the toroidal field component becomes non-monotonic for higher modes. 
}
\label{fig:lsxzeta}
\end{figure*}

We proceed now to analyze the influence of the magnetic field geometry on the frequencies. In Fig. \ref{fig:lsxzeta} we present results for a 1.4 $M_{\odot}$ SLy+SLy star in terms of $\zeta$, for a sequence of stars with fixed amplitude $10^{15}$ G of the magnetic field at the pole of the star. Our results show that the frequencies of the lower $l$ modes increase with  $\zeta$. This result can be explained based on the behavior shown in the lower plot Fig. \ref{fig:tests}, that shows how the frequencies grow with increasing magnetic field, if we note that along the sequence of stars shown in Fig. \ref{fig:lsxzeta} the total magnetic energy is also increasing due to the addition of the toroidal field component. (However, we will show in Sec. \ref{sec:evolution} that this effect persists for sequences with a fixed magnetic field energy.) The fundamental $l = 2$ mode is the most sensitive to $\zeta$, presenting a variation of $\approx 50\%$ in our range of $\zeta$. The variation in the frequencies decreases with increasing $l$, and we also note that for $l > 7$ the behavior is non-monotonic, but in these cases the variation is less than 2\%. Although the perturbation equation, given by eqs. (\ref{eq:Y})-(\ref{eq:coeffC}) is fairly involved, it is possible to provide an approximate explanation for this behavior. The $\zeta$-dependence of the torsional oscillations is embedded in the solution for the magnetic vector potential given by $a_1$, while the $l$-dependence is encoded in the parameters $\lambda$ and $\lambda_1$ (which is always multiplied by $a_1$ in eqs. (\ref{eq:coeffA})-(\ref{eq:coeffC})). It is easy to show that $\lambda_1 \approx -\frac{1}{4}(1 + 3/4l^2)$, which tends quickly to a constant value with increasing $l$. For $l = 7$, the correction to $\lambda_1$ is indeed approximately 2\%.

\begin{figure}
\begin{center}
\includegraphics[width=0.95\linewidth]{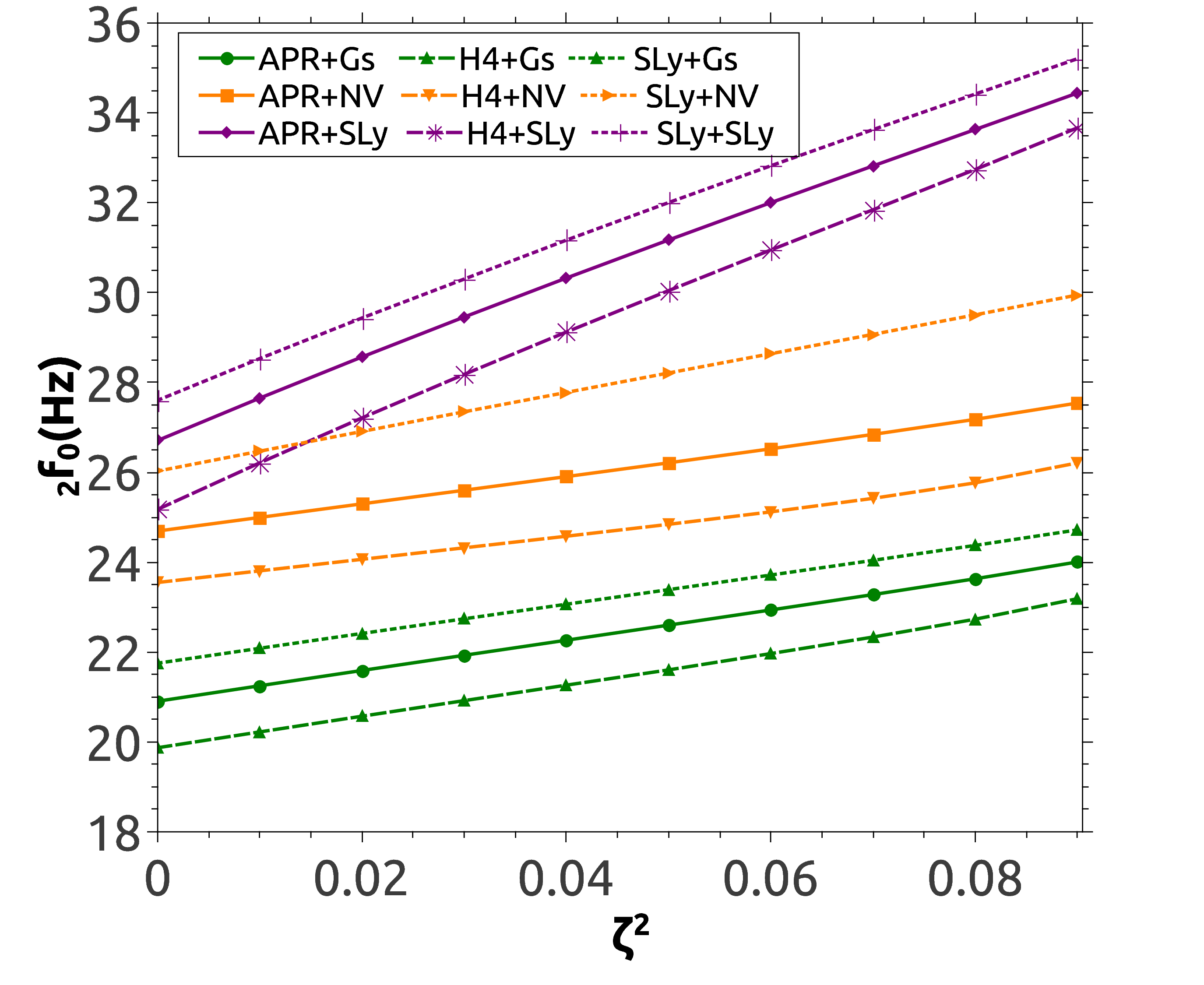}
\end{center}
\caption{Same as Fig. \ref{fig:lsxzeta}, but only for the $l = 2$ mode and for our set of 9 combinations of core+crust EOSs. The crust EOS is the most relevant for the behavior of the frequencies of the torsional oscillations. 
}
\label{fig:Bcte}
\end{figure}

Next we use our full set of 9 combinations of core+crust EOSs to explore the behavior of the fundamental mode, and the results are reported in Fig. \ref{fig:Bcte}. We can see now that the crust EOS is crucial for determining the values of the frequencies. The SLy crust is the most sensitive to the variation of the magnetic field geometry, and the frequency values increase by $\approx 30\%$ in comparison with the purely dipolar case, while the variations in the case of the Gs and NV crusts are $\approx 15\%$. This difference is caused by the higher values of the shear modulus for the SLy EOS, as quoted in Fig. \ref{fig:shear}.

We also compared our results for mixed poloidal-toroidal magnetic field configurations presented in Fig. \ref{fig:Bcte} with other works. First we note that a similar analysis was presented in the appendix of \cite{2008MNRAS.385.2161S} for a 1.4 $M_{\odot}$ polytropic star with a simple approximate formula for the shear modulus (with a typical value for the shear speed assumed constant inside the star). Although the results are strongly dependent on the choice of EOS for the core and crust and therefore only an approximate comparison with our models is possible, their results show the same upward trend for the frequencies with increasing $\zeta$ that we found. In particular, their results agree within less then 10\% with our H4+NV star presented in Fig. \ref{fig:Bcte}.

Another comparison can be performed with the results for mixed poloidal-toroidal field presented by \cite{2013MNRAS.430.1811G}, where the QPOs are expected at the turning points in the spectra of magneto-elastic oscillations. However, in their $B_x$ models the toroidal magnetic field component is restricted to the region of closed dipolar field lines inside the star, presenting a configuration that is very different from our model. Despite that, the results presented in their Fig. 8 for models with increasing poloidal and toroidal fields have turning points in the range $20-40$ Hz, in approximate agreement with our Fig. \ref{fig:Bcte}.

We note that even if we fix the mass of the star and magnetic field at the pole, as we did in Fig. \ref{fig:Bcte}, the same frequency value can correspond to different EOSs with different magnetic field configurations. These results add to those presented in Fig. \ref{fig:tests} to make it more difficult to solve the inverse problem, that is, to obtain the parameters of the star from its asteroseismology.

\begin{figure}
\begin{center}
\includegraphics[width=0.95\linewidth]{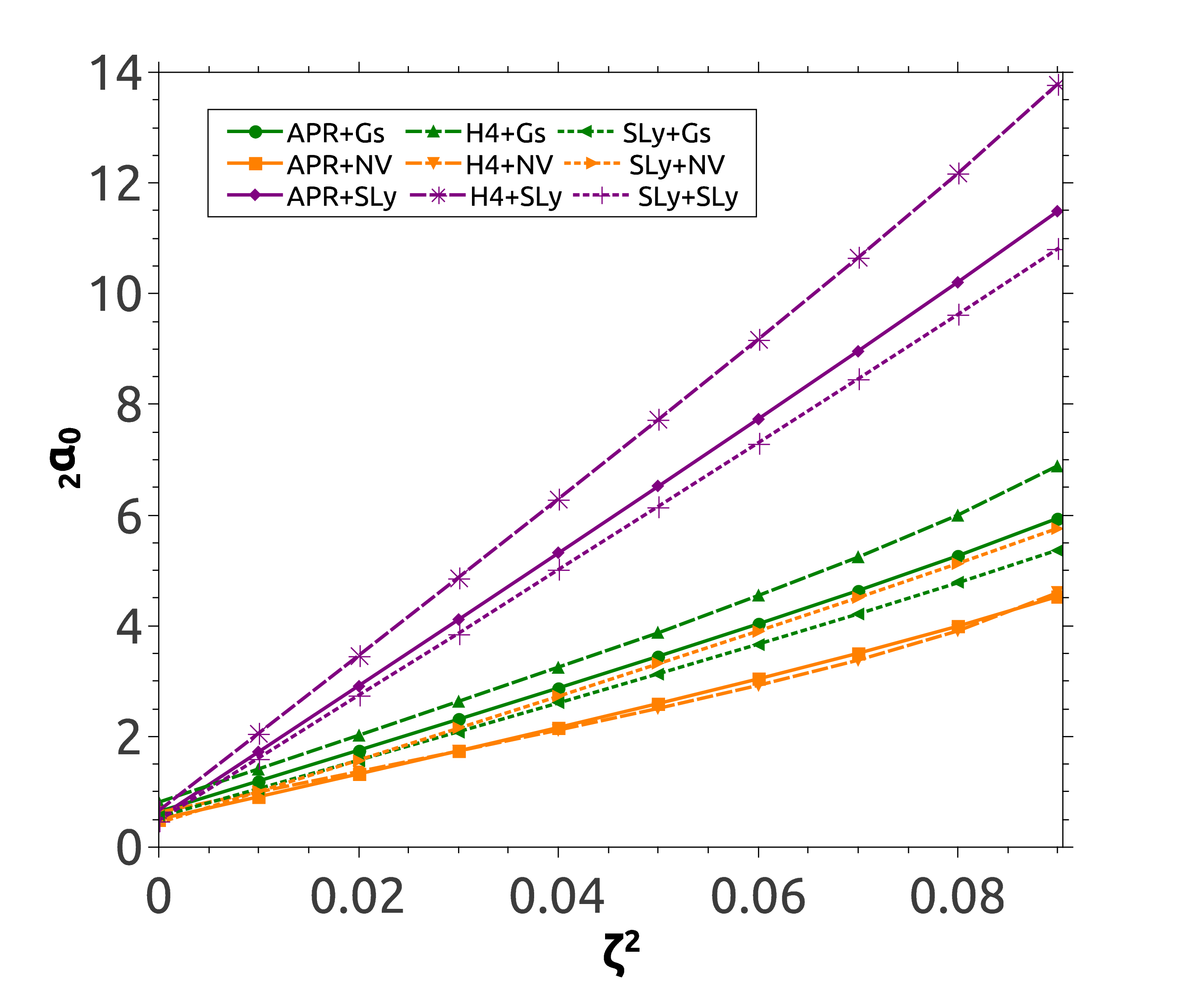}
\end{center}
\caption{Behaviour of the coefficient $_2\alpha_0$ of the magnetic field correction to the fundamental torsional frequency, defined in eq. (\ref{eq:f}), as a function of the magnetic field geometry as parametrized by $\zeta$. }
\label{fig:alphas}
\end{figure}

\begin{figure}
\begin{center}
\includegraphics[width=0.95\linewidth]{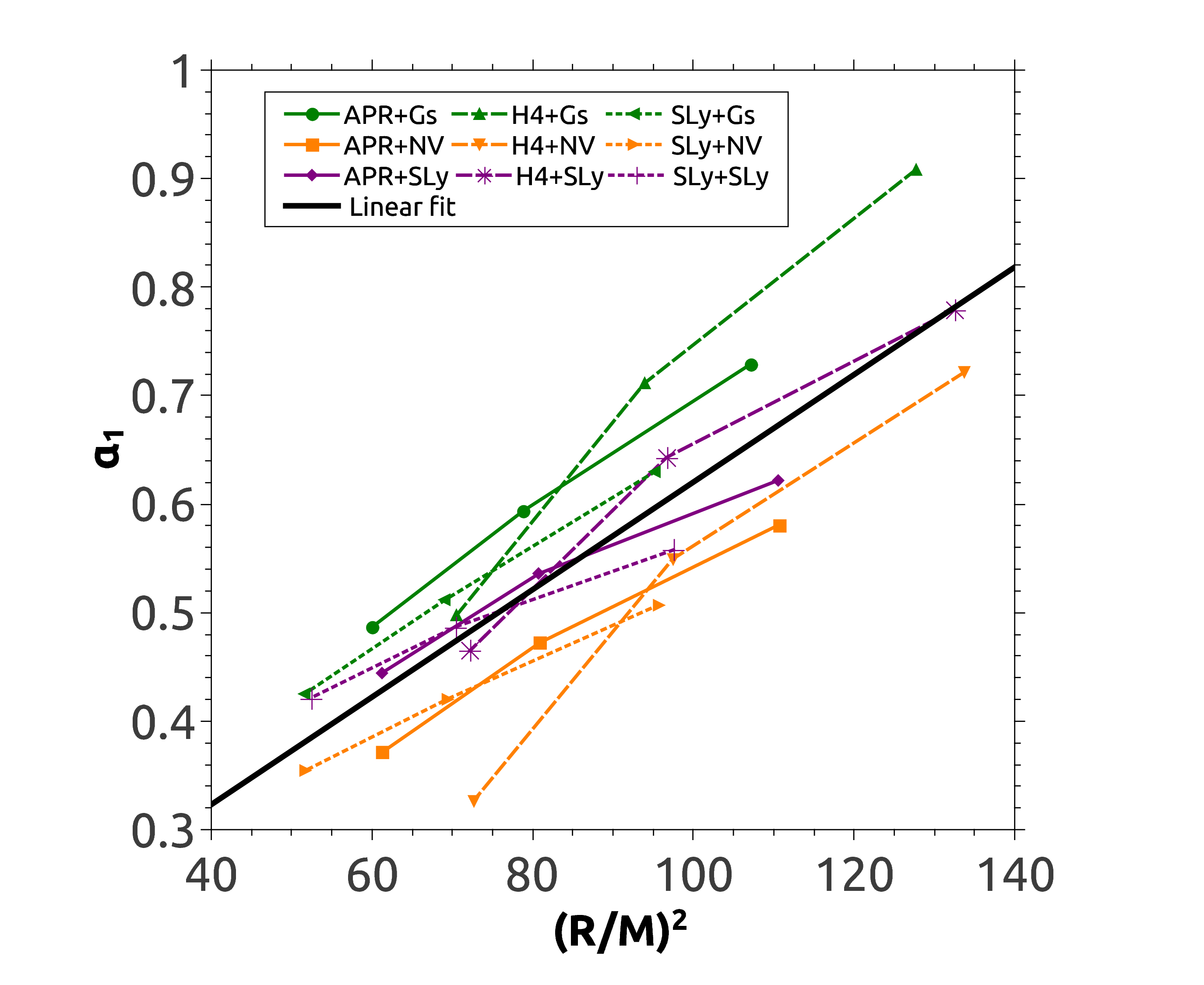}
\includegraphics[width=0.95\linewidth]{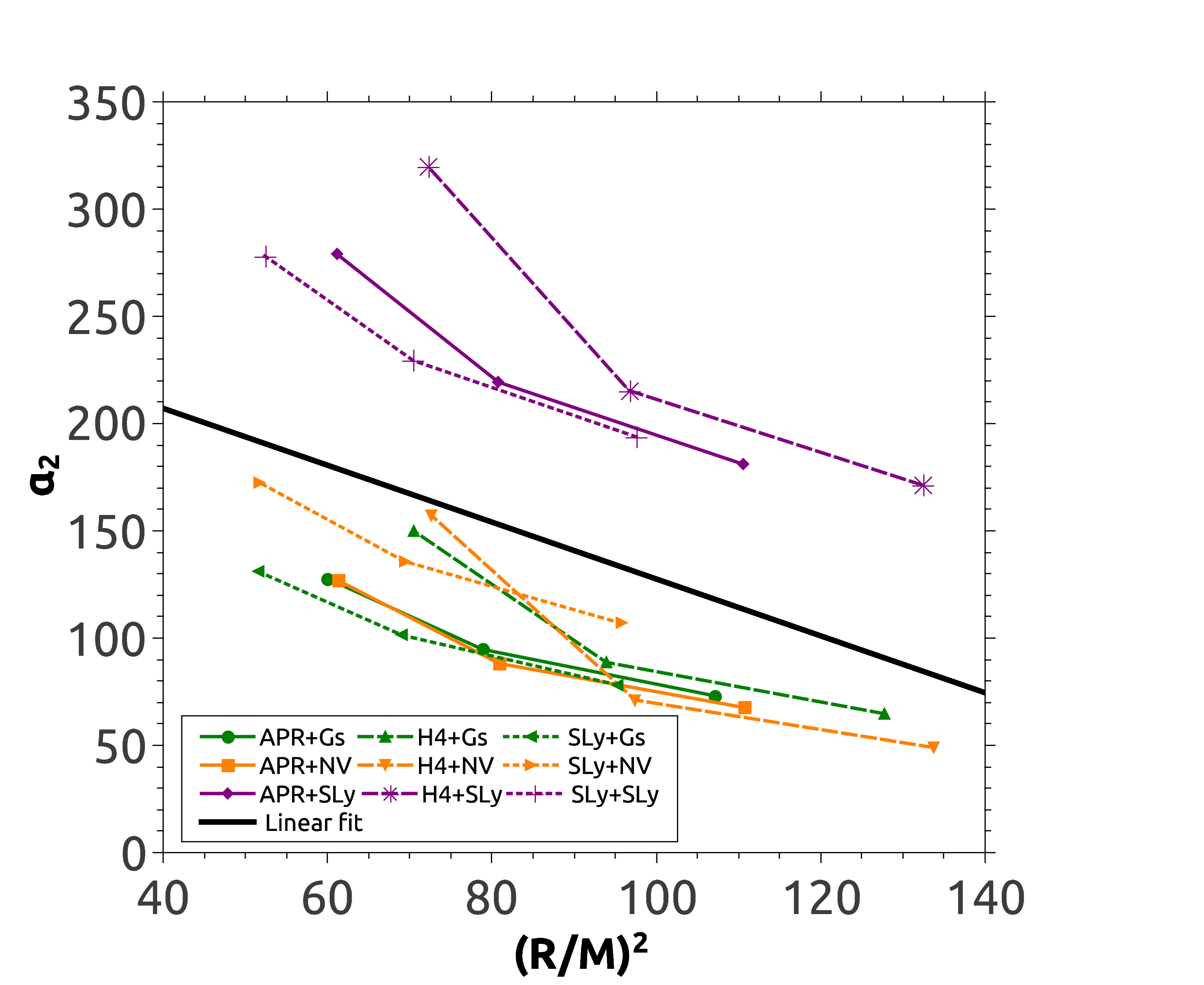}
\end{center}
\caption{Coefficients $\alpha_1$ and $\alpha_2$, defined in eq.(\ref{eq:alpha}), as functions of the stellar parameters. For the case of a purely dipolar field, the correction to the torsional frequency is entirely due to $\alpha_1$. The correlation found for $\alpha_1$ is close to EOS-independent, but the spread is larger in the case of $\alpha_2$.}
\label{fig:alpha12}
\end{figure}

In order to look for possible EOS-independent relations for these frequencies, we explored the way in which eq. (\ref{eq:f}) is modified by an increasing toroidal field component. We found that the parameter $\phantom{ }_2\alpha_0$ is no longer a constant for a given stellar mass and EOS, but it is now a function of $\zeta$. As we report in Fig. \ref{fig:alphas} for the same stellar sequences presented in Fig. \ref{fig:Bcte}, $\phantom{ }_2\alpha_0$ increases approximately linearly with $\zeta^2$.
Therefore we propose a simple parametrization of the form
\begin{equation}
\phantom{ }_2\alpha_0 = \alpha_1(1+ \alpha_2\zeta^2)\,,
\label{eq:alpha}
\end{equation} 
where for a pure dipole, $\phantom{ }_2\alpha_0 = \alpha_1$, whereas $\alpha_2$ gives the coefficient of the correction due to the toroidal field component. We explored their dependence on the stellar parameters and looked for any noticeable trends, motivated by some universal (EOS-independent) relations known for neutron star perturbations (see for instance \cite{2015PhRvD..91d4034C,2013PhRvD..88b3009Y}). By inspection we found that $\alpha_1$ increases linearly with the square of the reciprocal of the compactness $M/R$, and this behavior is approximately EOS-independent, see the upper plot of Fig. \ref{fig:alpha12}. For $\alpha_2$, we found the opposite behavior, although the spread due to the EOS is larger and mostly determined by the EOS of the crust, as can be seen in the lower plot of Fig. \ref{fig:alpha12}.

\section{Quasi-static evolution for the magnetic field}
\label{sec:evolution}

If the giant flare was caused by a rearrangement of the magnetic field of the star, it is plausible that it caused or was caused by a change in the magnetic field geometry. The frequencies of the QPOs could carry a signature of this process if their values are ``drifting" with time, and ideally could be used to track the time evolution of $\zeta$. In the analysis presented in Sec. \ref{sec:results}, we increased $\zeta$ while keeping the amplitude of the magnetic field at the pole fixed, which means that increasing the toroidal component resulted in an increase of the total magnetic energy inside the star. In order to simulate our evolution scenario without artificially increasing the magnetic energy inside the star, we kept it constant by lowering the magnetic field amplitude at the pole accordingly as we increased $\zeta$ in our next analysis.

\begin{figure}
\begin{center}
\includegraphics[width=0.95\linewidth]{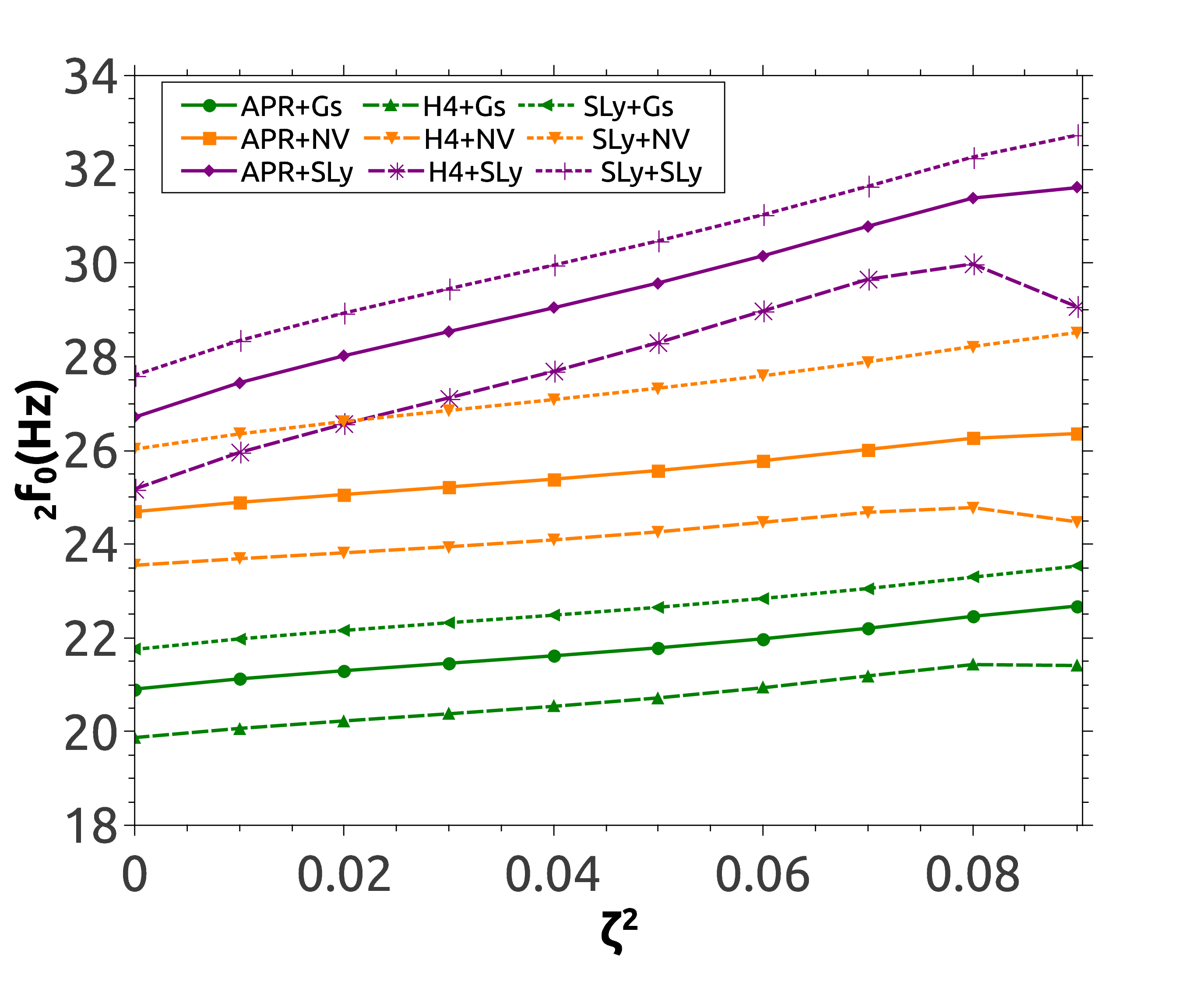}
\end{center}
\caption{Same as Fig. \ref{fig:Bcte}, but for sequences of stars with fixed total magnetic energy. The frequencies still increase for configurations with larger toroidal field contributions, but the fractional variation is approximately half of that in the case with a constant $B$ field at the pole of the star presented in Fig. \ref{fig:Bcte}.}
\label{fig:Ecte}
\end{figure}

Our results for this quasi-static evolution scenario are presented in Fig. \ref{fig:Ecte}. Comparing Figs. \ref{fig:Bcte} and \ref{fig:Ecte}, we see that the behavior is qualitatively similar: the frequencies increase with $\zeta$. However, this effect is less pronounced in \ref{fig:Ecte}, and the fractional variations are approximately half of those presented in Fig. \ref{fig:Bcte}. We believe that this is caused by the way in which the sequences of stars used in the two figures were constructed. In Fig. \ref{fig:Bcte}, the sequences of stars had constant magnetic field at the pole and an increasing toroidal component, which increased the total magnetic field energy, while in Fig. \ref{fig:Ecte} a fixed total magnetic energy is shared between the dipolar field and the increasing toroidal field. Therefore Fig. \ref{fig:Ecte} also shows that the field configuration alone is also responsible for changing the frequencies of the torsional modes.

Such an evolution of the magnetic field geometry could be responsible for a trend in the behavior of a long-lived QPO in the tail of a giant flare. If the field is rearranged from a more complicated mixed configuration (that could be caused by a persistent twisting of the field lines with the slow rotation of the magnetar) to a simpler pure dipole, the frequencies of the QPOs could go down in a noticeable way, if they can be determined with a resolution of 1 Hz or lower.

Lastly we performed an analysis of the energy in the perturbation, given by integral of the $\delta T^{tt}$ component of the linearized stress energy tensor over the volume of the crust. For the torsional modes of the crust, we have
\begin{equation}
\delta T^{tt} = 2H^{\alpha}\delta H_{\alpha}e^{-2\nu}\,,
\end{equation}
and after we substitute the magnetic field components given by eqs. (\ref{eq:B1})-(\ref{eq:B3}) (we remind that $H^{\alpha} = B^{\alpha}/4\pi$), and the perturbations $\delta H^{\alpha}$ in the magnetic field given by the induction equations (\ref{eq:induction}) in terms of $\delta u^{\phi}$, given by eq. (\ref{eq:delta_u_phi}), we obtain the total energy of the mode with the volume integral
\begin{equation}
\int_V \delta T^{tt}dV = \frac{8}{5}\sqrt{\pi}\zeta\int_{R_{\rm c}}^R e^{-3\nu}
a_1\left[a_1{\cal Y}_{,r} + 2a_{1,r}{\cal Y}\right]
e^{i\omega t}dr,
\label{eq:deltaT}
\end{equation}
where the only non-zero contribution comes from the $l = 2$ term. We remind here that our linear perturbation treatment of the modes does not allow us to make predictions on the amplitude of the mode. However, we can see that it depends linearly on the amplitude of the oscillation and quadratically on the magnetic field amplitude, as expected. We also stress that this lowest order contribution to the energy is present only in configurations with a toroidal field component and goes to zero as $\zeta \to 0$.

\begin{figure}
\begin{center}
\includegraphics[width=0.95\linewidth]{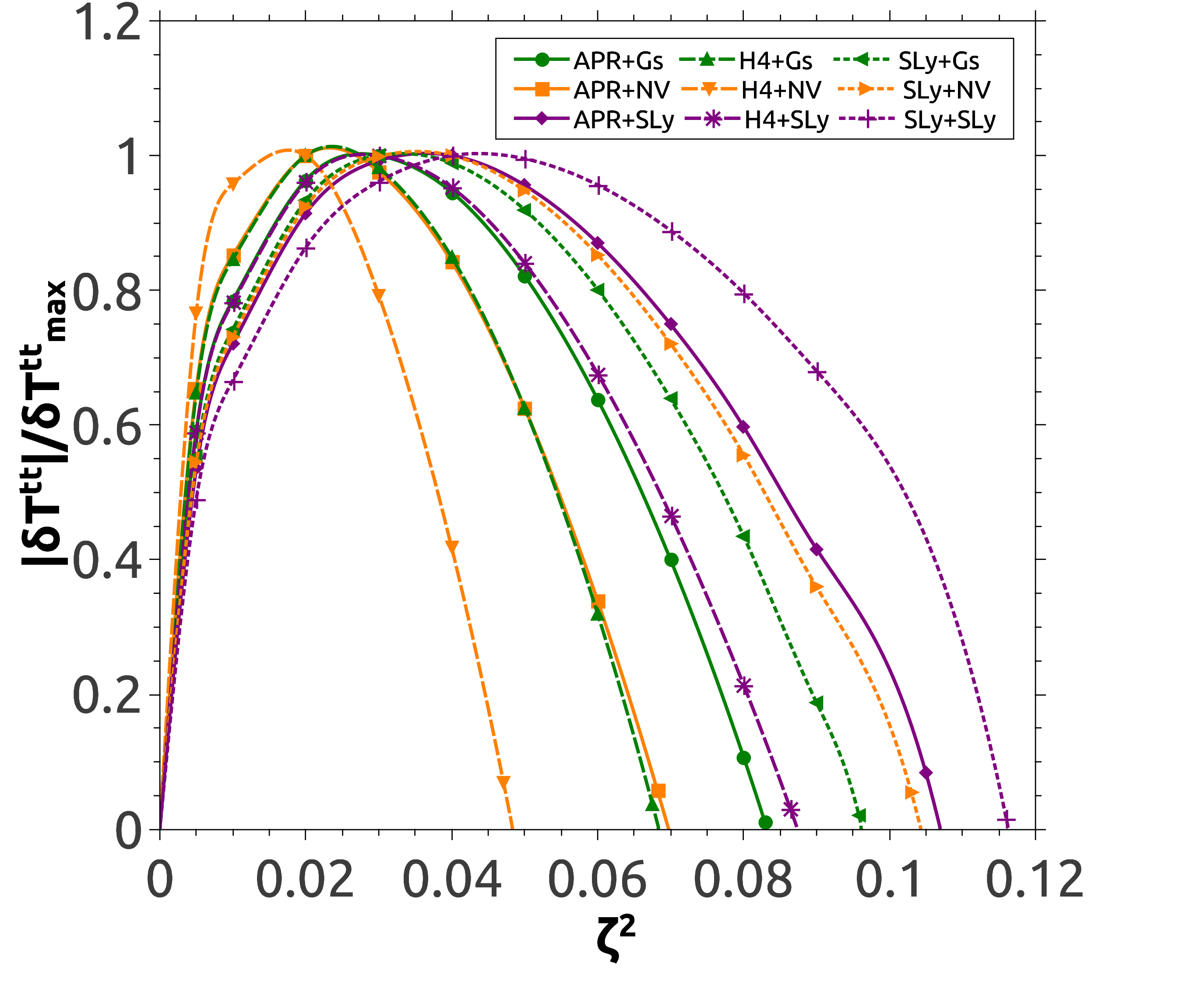}
\end{center}
\caption{Energy in the fundamental $l = 2$ mode integrated inside the stellar crust for the sequences of stars with fixed total magnetic energy presented in Fig. \ref{fig:Ecte}, normalized by the configuration with maximum energy for each core+crust EOS. The local maxima could be linked to the emission of the giant flare.}
\label{fig:energy}
\end{figure}

An unexpected result appears when we analyze the energy in the mode as a function of the magnetic field geometry, as we can see in Fig. \ref{fig:energy}. The change in the energy is non-monotonic with $\zeta$, and each combination of core+curst EOS has a magnetic field configuration that maximizes the energy at linear order in the amplitude of the mode. The results presented were obtained by numerically integrating eq. (\ref{eq:deltaT}) for the sequences of stars with fixed total magnetic energy presented in Fig. \ref{fig:Ecte}. In order to compare the perturbations obtained for different configurations, we kept the value of the amplitude of the oscillation constant at the base of the crust for all cases, and normalized the energy in the mode for each configuration in a given sequence by the maximum value obtained for that sequence.

Our results motivate us to propose a possible mechanism for the still unknown trigger of the magnetar giant flare. Small oscillations could be constantly (but not necessarily continuously), present in the crust as they could be frequently excited and re-excited by magnetic or crustal stresses, for example. If the magnetic field configuration of the star evolves over time, the energy in these modes of oscillation would peak for a certain contribution of the toroidal field component, at which point the flare could be triggered. A higher order analysis of this effect should be performed, of course, before any strong statements can be made. For instance, the increase in the energy of the modes should come from the magnetic field of the star, but the linear perturbation calculation does not include this backreaction. However, our analysis already points to a non-trivial behavior of the energy as a function of $\zeta$, which should be further explored.

\section{Conclusions}
\label{sec:conclusions}

The interiors of neutron stars, from their crusts to their cores, represent states of matter that cannot be accessed on
Earth. Their study therefore allows us to glimpse physics in a realm that constructively challenges our understanding
of nuclear and condensed matter physics. Since the first report of QPOs from SGR giant flares it has been thought that a
full understanding of the QPOs could give us unique insight into this mysterious realm. There are also some hopes that these giant flares, and possibly also the QPOs might eventually be detectable as gravitational wave events \cite{2011PhRvD..83j4014C,2017CQGra..34p4002Q}, but see also \citet{2011MNRAS.418..659L}, perhaps by third-generation detectors such as the \emph{Einstein Telescope} \cite{2010CQGra..27s4002P} or the \emph{Cosmic Explorer} \cite{2017CQGra..34d4001A}.

We have presented here a careful analysis of the frequencies of torsional modes of oscillation of the crust for magnetized neutron stars. Motivated by the observations of QPOs in the tail of the observed giant flares of SGRs and the prospect of doing neutron star asteroseismology, we explored the influence of the magnetic field geometry, encoded in the variable $\zeta$, in the values of the frequencies. We found that the frequency of the fundamental $l = 2$ torsional mode can change by up to $\approx 30\%$ with the variation of the magnetic field geometry. This variation would be detectable if the frequencies can be determined with a resolution of 1 Hz or lower, and could point to an overall evolution of the magnetic field geometry.

The solution of the inverse problem, that is, the calculation of the parameters of the star from its observed frequencies of oscillation, becomes complicated by the multidimensionality of the parameter space in this case. However, we were able to make considerable progress towards an approximately EOS-independent relation for the frequency of the $l = 2$ fundamental mode in the case when the magnetic field is (close to) a pure dipole. 

Finally, we examined the energy in the mode and found a dominant contribution in the linear order, that appears only in magnetic fields with a toroidal component. This energy presents a maximum for a specific (EOS-dependent) magnetic field configuration, which could be linked to the still unknown mechanism behind the triggering of the giant flare.

\section*{Acknowledgements}
We thank Hajime Sotani for sharing the A EOS table, which we used in our initial code tests, and Andrew Steiner for providing the Gs and SLy crust EOS tables used in this work. We also thank Kostas Kokkotas, Yuri Levin and Cole Miller for useful comments and suggestions. This work was supported by in part by the National Science Foundation under Grant No. NSF PHY-1748958, the Brazilian agency CAPES, the Brazilian National Council for Scientific and Technological Development (CNPq grant 303750/2017-0) and 
the S\~ao Paulo Research Foundation (FAPESP Grant 2015/20433-4). 

\bibliography{magnetic} 

\begin{thebibliography}{69}%
\makeatletter
\providecommand \@ifxundefined [1]{%
 \@ifx{#1\undefined}
}%
\providecommand \@ifnum [1]{%
 \ifnum #1\expandafter \@firstoftwo
 \else \expandafter \@secondoftwo
 \fi
}%
\providecommand \@ifx [1]{%
 \ifx #1\expandafter \@firstoftwo
 \else \expandafter \@secondoftwo
 \fi
}%
\providecommand \natexlab [1]{#1}%
\providecommand \enquote  [1]{``#1''}%
\providecommand \bibnamefont  [1]{#1}%
\providecommand \bibfnamefont [1]{#1}%
\providecommand \citenamefont [1]{#1}%
\providecommand \href@noop [0]{\@secondoftwo}%
\providecommand \href [0]{\begingroup \@sanitize@url \@href}%
\providecommand \@href[1]{\@@startlink{#1}\@@href}%
\providecommand \@@href[1]{\endgroup#1\@@endlink}%
\providecommand \@sanitize@url [0]{\catcode `\\12\catcode `\$12\catcode
  `\&12\catcode `\#12\catcode `\^12\catcode `\_12\catcode `\%12\relax}%
\providecommand \@@startlink[1]{}%
\providecommand \@@endlink[0]{}%
\providecommand \url  [0]{\begingroup\@sanitize@url \@url }%
\providecommand \@url [1]{\endgroup\@href {#1}{\urlprefix }}%
\providecommand \urlprefix  [0]{URL }%
\providecommand \Eprint [0]{\href }%
\providecommand \doibase [0]{http://dx.doi.org/}%
\providecommand \selectlanguage [0]{\@gobble}%
\providecommand \bibinfo  [0]{\@secondoftwo}%
\providecommand \bibfield  [0]{\@secondoftwo}%
\providecommand \translation [1]{[#1]}%
\providecommand \BibitemOpen [0]{}%
\providecommand \bibitemStop [0]{}%
\providecommand \bibitemNoStop [0]{.\EOS\space}%
\providecommand \EOS [0]{\spacefactor3000\relax}%
\providecommand \BibitemShut  [1]{\csname bibitem#1\endcsname}%
\let\auto@bib@innerbib\@empty
\bibitem [{\citenamefont {{Turolla}}\ \emph {et~al.}(2015)\citenamefont
  {{Turolla}}, \citenamefont {{Zane}},\ and\ \citenamefont
  {{Watts}}}]{2015RPPh...78k6901T}%
  \BibitemOpen
  \bibfield  {author} {\bibinfo {author} {\bibfnamefont {R.}~\bibnamefont
  {{Turolla}}}, \bibinfo {author} {\bibfnamefont {S.}~\bibnamefont {{Zane}}}, \
  and\ \bibinfo {author} {\bibfnamefont {A.~L.}\ \bibnamefont {{Watts}}},\
  }\bibfield  {title} {\enquote {\bibinfo {title} {{Magnetars: the physics
  behind observations. A review}},}\ }\href {\doibase
  10.1088/0034-4885/78/11/116901} {\bibfield  {journal} {\bibinfo  {journal}
  {Reports on Progress in Physics}\ }\textbf {\bibinfo {volume} {78}},\
  \bibinfo {eid} {116901} (\bibinfo {year} {2015})},\ \Eprint
  {http://arxiv.org/abs/1507.02924} {arXiv:1507.02924 [astro-ph.HE]}
  \BibitemShut {NoStop}%
\bibitem [{\citenamefont {{Duncan}}\ and\ \citenamefont
  {{Thompson}}(1992)}]{1992ApJ...392L...9D}%
  \BibitemOpen
  \bibfield  {author} {\bibinfo {author} {\bibfnamefont {R.~C.}\ \bibnamefont
  {{Duncan}}}\ and\ \bibinfo {author} {\bibfnamefont {C.}~\bibnamefont
  {{Thompson}}},\ }\bibfield  {title} {\enquote {\bibinfo {title} {{Formation
  of very strongly magnetized neutron stars - Implications for gamma-ray
  bursts}},}\ }\href {\doibase 10.1086/186413} {\bibfield  {journal} {\bibinfo
  {journal} {\apjl}\ }\textbf {\bibinfo {volume} {392}},\ \bibinfo {pages}
  {L9--L13} (\bibinfo {year} {1992})}\BibitemShut {NoStop}%
\bibitem [{\citenamefont {{Thompson}}\ and\ \citenamefont
  {{Duncan}}(1995)}]{1995MNRAS.275..255T}%
  \BibitemOpen
  \bibfield  {author} {\bibinfo {author} {\bibfnamefont {C.}~\bibnamefont
  {{Thompson}}}\ and\ \bibinfo {author} {\bibfnamefont {R.~C.}\ \bibnamefont
  {{Duncan}}},\ }\bibfield  {title} {\enquote {\bibinfo {title} {{The soft
  gamma repeaters as very strongly magnetized neutron stars - I. Radiative
  mechanism for outbursts}},}\ }\href {\doibase 10.1093/mnras/275.2.255}
  {\bibfield  {journal} {\bibinfo  {journal} {\mnras}\ }\textbf {\bibinfo
  {volume} {275}},\ \bibinfo {pages} {255--300} (\bibinfo {year}
  {1995})}\BibitemShut {NoStop}%
\bibitem [{\citenamefont {{Mazets}}\ \emph {et~al.}(1979)\citenamefont
  {{Mazets}}, \citenamefont {{Golentskii}}, \citenamefont {{Ilinskii}},
  \citenamefont {{Aptekar}},\ and\ \citenamefont
  {{Guryan}}}]{1979Natur.282..587M}%
  \BibitemOpen
  \bibfield  {author} {\bibinfo {author} {\bibfnamefont {E.~P.}\ \bibnamefont
  {{Mazets}}}, \bibinfo {author} {\bibfnamefont {S.~V.}\ \bibnamefont
  {{Golentskii}}}, \bibinfo {author} {\bibfnamefont {V.~N.}\ \bibnamefont
  {{Ilinskii}}}, \bibinfo {author} {\bibfnamefont {R.~L.}\ \bibnamefont
  {{Aptekar}}}, \ and\ \bibinfo {author} {\bibfnamefont {I.~A.}\ \bibnamefont
  {{Guryan}}},\ }\bibfield  {title} {\enquote {\bibinfo {title} {{Observations
  of a flaring X-ray pulsar in Dorado}},}\ }\href {\doibase 10.1038/282587a0}
  {\bibfield  {journal} {\bibinfo  {journal} {\nat}\ }\textbf {\bibinfo
  {volume} {282}},\ \bibinfo {pages} {587--589} (\bibinfo {year}
  {1979})}\BibitemShut {NoStop}%
\bibitem [{\citenamefont {{Barat}}\ \emph {et~al.}(1983)\citenamefont
  {{Barat}}, \citenamefont {{Hayles}}, \citenamefont {{Hurley}}, \citenamefont
  {{Niel}}, \citenamefont {{Vedrenne}}, \citenamefont {{Desai}}, \citenamefont
  {{Kurt}}, \citenamefont {{Zenchenko}},\ and\ \citenamefont
  {{Estulin}}}]{1983A&A...126..400B}%
  \BibitemOpen
  \bibfield  {author} {\bibinfo {author} {\bibfnamefont {C.}~\bibnamefont
  {{Barat}}}, \bibinfo {author} {\bibfnamefont {R.~I.}\ \bibnamefont
  {{Hayles}}}, \bibinfo {author} {\bibfnamefont {K.}~\bibnamefont {{Hurley}}},
  \bibinfo {author} {\bibfnamefont {M.}~\bibnamefont {{Niel}}}, \bibinfo
  {author} {\bibfnamefont {G.}~\bibnamefont {{Vedrenne}}}, \bibinfo {author}
  {\bibfnamefont {U.}~\bibnamefont {{Desai}}}, \bibinfo {author} {\bibfnamefont
  {V.~G.}\ \bibnamefont {{Kurt}}}, \bibinfo {author} {\bibfnamefont {V.~M.}\
  \bibnamefont {{Zenchenko}}}, \ and\ \bibinfo {author} {\bibfnamefont {I.~V.}\
  \bibnamefont {{Estulin}}},\ }\bibfield  {title} {\enquote {\bibinfo {title}
  {{Fine time structure in the 1979 March 5 gamma ray burst}},}\ }\href@noop {}
  {\bibfield  {journal} {\bibinfo  {journal} {\aap}\ }\textbf {\bibinfo
  {volume} {126}},\ \bibinfo {pages} {400--402} (\bibinfo {year}
  {1983})}\BibitemShut {NoStop}%
\bibitem [{\citenamefont {{Hurley}}\ \emph {et~al.}(1999)\citenamefont
  {{Hurley}}, \citenamefont {{Cline}}, \citenamefont {{Mazets}}, \citenamefont
  {{Barthelmy}}, \citenamefont {{Butterworth}}, \citenamefont {{Marshall}},
  \citenamefont {{Palmer}}, \citenamefont {{Aptekar}}, \citenamefont
  {{Golenetskii}}, \citenamefont {{Il'Inskii}}, \citenamefont {{Frederiks}},
  \citenamefont {{McTiernan}}, \citenamefont {{Gold}},\ and\ \citenamefont
  {{Trombka}}}]{1999Natur.397...41H}%
  \BibitemOpen
  \bibfield  {author} {\bibinfo {author} {\bibfnamefont {K.}~\bibnamefont
  {{Hurley}}}, \bibinfo {author} {\bibfnamefont {T.}~\bibnamefont {{Cline}}},
  \bibinfo {author} {\bibfnamefont {E.}~\bibnamefont {{Mazets}}}, \bibinfo
  {author} {\bibfnamefont {S.}~\bibnamefont {{Barthelmy}}}, \bibinfo {author}
  {\bibfnamefont {P.}~\bibnamefont {{Butterworth}}}, \bibinfo {author}
  {\bibfnamefont {F.}~\bibnamefont {{Marshall}}}, \bibinfo {author}
  {\bibfnamefont {D.}~\bibnamefont {{Palmer}}}, \bibinfo {author}
  {\bibfnamefont {R.}~\bibnamefont {{Aptekar}}}, \bibinfo {author}
  {\bibfnamefont {S.}~\bibnamefont {{Golenetskii}}}, \bibinfo {author}
  {\bibfnamefont {V.}~\bibnamefont {{Il'Inskii}}}, \bibinfo {author}
  {\bibfnamefont {D.}~\bibnamefont {{Frederiks}}}, \bibinfo {author}
  {\bibfnamefont {J.}~\bibnamefont {{McTiernan}}}, \bibinfo {author}
  {\bibfnamefont {R.}~\bibnamefont {{Gold}}}, \ and\ \bibinfo {author}
  {\bibfnamefont {J.}~\bibnamefont {{Trombka}}},\ }\bibfield  {title} {\enquote
  {\bibinfo {title} {{A giant periodic flare from the soft {$\gamma$}-ray
  repeater SGR1900+14}},}\ }\href {\doibase 10.1038/16199} {\bibfield
  {journal} {\bibinfo  {journal} {\nat}\ }\textbf {\bibinfo {volume} {397}},\
  \bibinfo {pages} {41--43} (\bibinfo {year} {1999})},\ \Eprint
  {http://arxiv.org/abs/astro-ph/9811443} {astro-ph/9811443} \BibitemShut
  {NoStop}%
\bibitem [{\citenamefont {{Terasawa}}\ \emph {et~al.}(2005)\citenamefont
  {{Terasawa}}, \citenamefont {{Tanaka}}, \citenamefont {{Takei}},
  \citenamefont {{Kawai}}, \citenamefont {{Yoshida}}, \citenamefont {{Nomoto}},
  \citenamefont {{Yoshikawa}}, \citenamefont {{Saito}}, \citenamefont
  {{Kasaba}}, \citenamefont {{Takashima}}, \citenamefont {{Mukai}},
  \citenamefont {{Noda}}, \citenamefont {{Murakami}}, \citenamefont
  {{Watanabe}}, \citenamefont {{Muraki}}, \citenamefont {{Yokoyama}},\ and\
  \citenamefont {{Hoshino}}}]{2005Natur.434.1110T}%
  \BibitemOpen
  \bibfield  {author} {\bibinfo {author} {\bibfnamefont {T.}~\bibnamefont
  {{Terasawa}}}, \bibinfo {author} {\bibfnamefont {Y.~T.}\ \bibnamefont
  {{Tanaka}}}, \bibinfo {author} {\bibfnamefont {Y.}~\bibnamefont {{Takei}}},
  \bibinfo {author} {\bibfnamefont {N.}~\bibnamefont {{Kawai}}}, \bibinfo
  {author} {\bibfnamefont {A.}~\bibnamefont {{Yoshida}}}, \bibinfo {author}
  {\bibfnamefont {K.}~\bibnamefont {{Nomoto}}}, \bibinfo {author}
  {\bibfnamefont {I.}~\bibnamefont {{Yoshikawa}}}, \bibinfo {author}
  {\bibfnamefont {Y.}~\bibnamefont {{Saito}}}, \bibinfo {author} {\bibfnamefont
  {Y.}~\bibnamefont {{Kasaba}}}, \bibinfo {author} {\bibfnamefont
  {T.}~\bibnamefont {{Takashima}}}, \bibinfo {author} {\bibfnamefont
  {T.}~\bibnamefont {{Mukai}}}, \bibinfo {author} {\bibfnamefont
  {H.}~\bibnamefont {{Noda}}}, \bibinfo {author} {\bibfnamefont
  {T.}~\bibnamefont {{Murakami}}}, \bibinfo {author} {\bibfnamefont
  {K.}~\bibnamefont {{Watanabe}}}, \bibinfo {author} {\bibfnamefont
  {Y.}~\bibnamefont {{Muraki}}}, \bibinfo {author} {\bibfnamefont
  {T.}~\bibnamefont {{Yokoyama}}}, \ and\ \bibinfo {author} {\bibfnamefont
  {M.}~\bibnamefont {{Hoshino}}},\ }\bibfield  {title} {\enquote {\bibinfo
  {title} {{Repeated injections of energy in the first 600ms of the giant flare
  of SGR1806 - 20}},}\ }\href {\doibase 10.1038/nature03573} {\bibfield
  {journal} {\bibinfo  {journal} {\nat}\ }\textbf {\bibinfo {volume} {434}},\
  \bibinfo {pages} {1110--1111} (\bibinfo {year} {2005})},\ \Eprint
  {http://arxiv.org/abs/astro-ph/0502315} {astro-ph/0502315} \BibitemShut
  {NoStop}%
\bibitem [{\citenamefont {{Palmer}}\ \emph {et~al.}(2005)\citenamefont
  {{Palmer}}, \citenamefont {{Barthelmy}}, \citenamefont {{Gehrels}},
  \citenamefont {{Kippen}}, \citenamefont {{Cayton}}, \citenamefont
  {{Kouveliotou}}, \citenamefont {{Eichler}}, \citenamefont {{Wijers}},
  \citenamefont {{Woods}}, \citenamefont {{Granot}}, \citenamefont
  {{Lyubarsky}}, \citenamefont {{Ramirez-Ruiz}}, \citenamefont {{Barbier}},
  \citenamefont {{Chester}}, \citenamefont {{Cummings}}, \citenamefont
  {{Fenimore}}, \citenamefont {{Finger}}, \citenamefont {{Gaensler}},
  \citenamefont {{Hullinger}}, \citenamefont {{Krimm}}, \citenamefont
  {{Markwardt}}, \citenamefont {{Nousek}}, \citenamefont {{Parsons}},
  \citenamefont {{Patel}}, \citenamefont {{Sakamoto}}, \citenamefont {{Sato}},
  \citenamefont {{Suzuki}},\ and\ \citenamefont
  {{Tueller}}}]{2005Natur.434.1107P}%
  \BibitemOpen
  \bibfield  {author} {\bibinfo {author} {\bibfnamefont {D.~M.}\ \bibnamefont
  {{Palmer}}}, \bibinfo {author} {\bibfnamefont {S.}~\bibnamefont
  {{Barthelmy}}}, \bibinfo {author} {\bibfnamefont {N.}~\bibnamefont
  {{Gehrels}}}, \bibinfo {author} {\bibfnamefont {R.~M.}\ \bibnamefont
  {{Kippen}}}, \bibinfo {author} {\bibfnamefont {T.}~\bibnamefont {{Cayton}}},
  \bibinfo {author} {\bibfnamefont {C.}~\bibnamefont {{Kouveliotou}}}, \bibinfo
  {author} {\bibfnamefont {D.}~\bibnamefont {{Eichler}}}, \bibinfo {author}
  {\bibfnamefont {R.~A.~M.~J.}\ \bibnamefont {{Wijers}}}, \bibinfo {author}
  {\bibfnamefont {P.~M.}\ \bibnamefont {{Woods}}}, \bibinfo {author}
  {\bibfnamefont {J.}~\bibnamefont {{Granot}}}, \bibinfo {author}
  {\bibfnamefont {Y.~E.}\ \bibnamefont {{Lyubarsky}}}, \bibinfo {author}
  {\bibfnamefont {E.}~\bibnamefont {{Ramirez-Ruiz}}}, \bibinfo {author}
  {\bibfnamefont {L.}~\bibnamefont {{Barbier}}}, \bibinfo {author}
  {\bibfnamefont {M.}~\bibnamefont {{Chester}}}, \bibinfo {author}
  {\bibfnamefont {J.}~\bibnamefont {{Cummings}}}, \bibinfo {author}
  {\bibfnamefont {E.~E.}\ \bibnamefont {{Fenimore}}}, \bibinfo {author}
  {\bibfnamefont {M.~H.}\ \bibnamefont {{Finger}}}, \bibinfo {author}
  {\bibfnamefont {B.~M.}\ \bibnamefont {{Gaensler}}}, \bibinfo {author}
  {\bibfnamefont {D.}~\bibnamefont {{Hullinger}}}, \bibinfo {author}
  {\bibfnamefont {H.}~\bibnamefont {{Krimm}}}, \bibinfo {author} {\bibfnamefont
  {C.~B.}\ \bibnamefont {{Markwardt}}}, \bibinfo {author} {\bibfnamefont
  {J.~A.}\ \bibnamefont {{Nousek}}}, \bibinfo {author} {\bibfnamefont
  {A.}~\bibnamefont {{Parsons}}}, \bibinfo {author} {\bibfnamefont
  {S.}~\bibnamefont {{Patel}}}, \bibinfo {author} {\bibfnamefont
  {T.}~\bibnamefont {{Sakamoto}}}, \bibinfo {author} {\bibfnamefont
  {G.}~\bibnamefont {{Sato}}}, \bibinfo {author} {\bibfnamefont
  {M.}~\bibnamefont {{Suzuki}}}, \ and\ \bibinfo {author} {\bibfnamefont
  {J.}~\bibnamefont {{Tueller}}},\ }\bibfield  {title} {\enquote {\bibinfo
  {title} {{A giant {$\gamma$}-ray flare from the magnetar SGR 1806 - 20}},}\
  }\href {\doibase 10.1038/nature03525} {\bibfield  {journal} {\bibinfo
  {journal} {\nat}\ }\textbf {\bibinfo {volume} {434}},\ \bibinfo {pages}
  {1107--1109} (\bibinfo {year} {2005})},\ \Eprint
  {http://arxiv.org/abs/astro-ph/0503030} {astro-ph/0503030} \BibitemShut
  {NoStop}%
\bibitem [{\citenamefont {{Strohmayer}}\ and\ \citenamefont
  {{Watts}}(2005)}]{2005ApJ...632L.111S}%
  \BibitemOpen
  \bibfield  {author} {\bibinfo {author} {\bibfnamefont {T.~E.}\ \bibnamefont
  {{Strohmayer}}}\ and\ \bibinfo {author} {\bibfnamefont {A.~L.}\ \bibnamefont
  {{Watts}}},\ }\bibfield  {title} {\enquote {\bibinfo {title} {{Discovery of
  Fast X-Ray Oscillations during the 1998 Giant Flare from SGR 1900+14}},}\
  }\href {\doibase 10.1086/497911} {\bibfield  {journal} {\bibinfo  {journal}
  {\apjl}\ }\textbf {\bibinfo {volume} {632}},\ \bibinfo {pages} {L111--L114}
  (\bibinfo {year} {2005})},\ \Eprint {http://arxiv.org/abs/astro-ph/0508206}
  {astro-ph/0508206} \BibitemShut {NoStop}%
\bibitem [{\citenamefont {{Israel}}\ \emph {et~al.}(2005)\citenamefont
  {{Israel}}, \citenamefont {{Belloni}}, \citenamefont {{Stella}},
  \citenamefont {{Rephaeli}}, \citenamefont {{Gruber}}, \citenamefont
  {{Casella}}, \citenamefont {{Dall'Osso}}, \citenamefont {{Rea}},
  \citenamefont {{Persic}},\ and\ \citenamefont
  {{Rothschild}}}]{2005ApJ...628L..53I}%
  \BibitemOpen
  \bibfield  {author} {\bibinfo {author} {\bibfnamefont {G.~L.}\ \bibnamefont
  {{Israel}}}, \bibinfo {author} {\bibfnamefont {T.}~\bibnamefont {{Belloni}}},
  \bibinfo {author} {\bibfnamefont {L.}~\bibnamefont {{Stella}}}, \bibinfo
  {author} {\bibfnamefont {Y.}~\bibnamefont {{Rephaeli}}}, \bibinfo {author}
  {\bibfnamefont {D.~E.}\ \bibnamefont {{Gruber}}}, \bibinfo {author}
  {\bibfnamefont {P.}~\bibnamefont {{Casella}}}, \bibinfo {author}
  {\bibfnamefont {S.}~\bibnamefont {{Dall'Osso}}}, \bibinfo {author}
  {\bibfnamefont {N.}~\bibnamefont {{Rea}}}, \bibinfo {author} {\bibfnamefont
  {M.}~\bibnamefont {{Persic}}}, \ and\ \bibinfo {author} {\bibfnamefont
  {R.~E.}\ \bibnamefont {{Rothschild}}},\ }\bibfield  {title} {\enquote
  {\bibinfo {title} {{The Discovery of Rapid X-Ray Oscillations in the Tail of
  the SGR 1806-20 Hyperflare}},}\ }\href {\doibase 10.1086/432615} {\bibfield
  {journal} {\bibinfo  {journal} {\apjl}\ }\textbf {\bibinfo {volume} {628}},\
  \bibinfo {pages} {L53--L56} (\bibinfo {year} {2005})},\ \Eprint
  {http://arxiv.org/abs/astro-ph/0505255} {astro-ph/0505255} \BibitemShut
  {NoStop}%
\bibitem [{\citenamefont {{Watts}}\ and\ \citenamefont
  {{Strohmayer}}(2006)}]{2006ApJ...637L.117W}%
  \BibitemOpen
  \bibfield  {author} {\bibinfo {author} {\bibfnamefont {A.~L.}\ \bibnamefont
  {{Watts}}}\ and\ \bibinfo {author} {\bibfnamefont {T.~E.}\ \bibnamefont
  {{Strohmayer}}},\ }\bibfield  {title} {\enquote {\bibinfo {title} {{Detection
  with RHESSI of High-Frequency X-Ray Oscillations in the Tailof the 2004
  Hyperflare from SGR 1806-20}},}\ }\href {\doibase 10.1086/500735} {\bibfield
  {journal} {\bibinfo  {journal} {\apjl}\ }\textbf {\bibinfo {volume} {637}},\
  \bibinfo {pages} {L117--L120} (\bibinfo {year} {2006})},\ \Eprint
  {http://arxiv.org/abs/astro-ph/0512630} {astro-ph/0512630} \BibitemShut
  {NoStop}%
\bibitem [{\citenamefont {{Strohmayer}}\ and\ \citenamefont
  {{Watts}}(2006)}]{2006ApJ...653..593S}%
  \BibitemOpen
  \bibfield  {author} {\bibinfo {author} {\bibfnamefont {T.~E.}\ \bibnamefont
  {{Strohmayer}}}\ and\ \bibinfo {author} {\bibfnamefont {A.~L.}\ \bibnamefont
  {{Watts}}},\ }\bibfield  {title} {\enquote {\bibinfo {title} {{The 2004
  Hyperflare from SGR 1806-20: Further Evidence for Global Torsional
  Vibrations}},}\ }\href {\doibase 10.1086/508703} {\bibfield  {journal}
  {\bibinfo  {journal} {\apj}\ }\textbf {\bibinfo {volume} {653}},\ \bibinfo
  {pages} {593--601} (\bibinfo {year} {2006})},\ \Eprint
  {http://arxiv.org/abs/astro-ph/0608463} {astro-ph/0608463} \BibitemShut
  {NoStop}%
\bibitem [{\citenamefont {{Huppenkothen}}\ \emph
  {et~al.}(2014{\natexlab{a}})\citenamefont {{Huppenkothen}}, \citenamefont
  {{Heil}}, \citenamefont {{Watts}},\ and\ \citenamefont {{G{\"o}{\u g}{\"u}{\c
  s}}}}]{2014ApJ...795..114H}%
  \BibitemOpen
  \bibfield  {author} {\bibinfo {author} {\bibfnamefont {D.}~\bibnamefont
  {{Huppenkothen}}}, \bibinfo {author} {\bibfnamefont {L.~M.}\ \bibnamefont
  {{Heil}}}, \bibinfo {author} {\bibfnamefont {A.~L.}\ \bibnamefont {{Watts}}},
  \ and\ \bibinfo {author} {\bibfnamefont {E.}~\bibnamefont {{G{\"o}{\u
  g}{\"u}{\c s}}}},\ }\bibfield  {title} {\enquote {\bibinfo {title}
  {{Quasi-periodic Oscillations in Short Recurring Bursts of Magnetars SGR
  1806-20 and SGR 1900+14 Observed with RXTE}},}\ }\href {\doibase
  10.1088/0004-637X/795/2/114} {\bibfield  {journal} {\bibinfo  {journal}
  {\apj}\ }\textbf {\bibinfo {volume} {795}},\ \bibinfo {eid} {114} (\bibinfo
  {year} {2014}{\natexlab{a}})},\ \Eprint {http://arxiv.org/abs/1409.7642}
  {arXiv:1409.7642 [astro-ph.HE]} \BibitemShut {NoStop}%
\bibitem [{\citenamefont {{Thompson}}\ and\ \citenamefont
  {{Duncan}}(1996)}]{1996ApJ...473..322T}%
  \BibitemOpen
  \bibfield  {author} {\bibinfo {author} {\bibfnamefont {C.}~\bibnamefont
  {{Thompson}}}\ and\ \bibinfo {author} {\bibfnamefont {R.~C.}\ \bibnamefont
  {{Duncan}}},\ }\bibfield  {title} {\enquote {\bibinfo {title} {{The Soft
  Gamma Repeaters as Very Strongly Magnetized Neutron Stars. II. Quiescent
  Neutrino, X-Ray, and Alfven Wave Emission}},}\ }\href {\doibase
  10.1086/178147} {\bibfield  {journal} {\bibinfo  {journal} {\apj}\ }\textbf
  {\bibinfo {volume} {473}},\ \bibinfo {pages} {322} (\bibinfo {year}
  {1996})}\BibitemShut {NoStop}%
\bibitem [{\citenamefont {{Lyutikov}}(2003)}]{2003MNRAS.346..540L}%
  \BibitemOpen
  \bibfield  {author} {\bibinfo {author} {\bibfnamefont {M.}~\bibnamefont
  {{Lyutikov}}},\ }\bibfield  {title} {\enquote {\bibinfo {title} {{Explosive
  reconnection in magnetars}},}\ }\href {\doibase
  10.1046/j.1365-2966.2003.07110.x} {\bibfield  {journal} {\bibinfo  {journal}
  {\mnras}\ }\textbf {\bibinfo {volume} {346}},\ \bibinfo {pages} {540--554}
  (\bibinfo {year} {2003})},\ \Eprint {http://arxiv.org/abs/astro-ph/0303384}
  {astro-ph/0303384} \BibitemShut {NoStop}%
\bibitem [{\citenamefont {{Timokhin}}\ \emph {et~al.}(2008)\citenamefont
  {{Timokhin}}, \citenamefont {{Eichler}},\ and\ \citenamefont
  {{Lyubarsky}}}]{2008ApJ...680.1398T}%
  \BibitemOpen
  \bibfield  {author} {\bibinfo {author} {\bibfnamefont {A.~N.}\ \bibnamefont
  {{Timokhin}}}, \bibinfo {author} {\bibfnamefont {D.}~\bibnamefont
  {{Eichler}}}, \ and\ \bibinfo {author} {\bibfnamefont {Y.}~\bibnamefont
  {{Lyubarsky}}},\ }\bibfield  {title} {\enquote {\bibinfo {title} {{On the
  Nature of Quasi-periodic Oscillations in the Tail of Soft Gamma Repeater
  Giant Flares}},}\ }\href {\doibase 10.1086/587925} {\bibfield  {journal}
  {\bibinfo  {journal} {\apj}\ }\textbf {\bibinfo {volume} {680}},\ \bibinfo
  {eid} {1398-1404} (\bibinfo {year} {2008})},\ \Eprint
  {http://arxiv.org/abs/0706.3698} {arXiv:0706.3698} \BibitemShut {NoStop}%
\bibitem [{\citenamefont {{Duncan}}(1998)}]{1998ApJ...498L..45D}%
  \BibitemOpen
  \bibfield  {author} {\bibinfo {author} {\bibfnamefont {R.~C.}\ \bibnamefont
  {{Duncan}}},\ }\bibfield  {title} {\enquote {\bibinfo {title} {{Global
  Seismic Oscillations in Soft Gamma Repeaters}},}\ }\href {\doibase
  10.1086/311303} {\bibfield  {journal} {\bibinfo  {journal} {\apjl}\ }\textbf
  {\bibinfo {volume} {498}},\ \bibinfo {pages} {L45--L49} (\bibinfo {year}
  {1998})},\ \Eprint {http://arxiv.org/abs/astro-ph/9803060} {astro-ph/9803060}
  \BibitemShut {NoStop}%
\bibitem [{\citenamefont {{Levin}}(2006)}]{2006MNRAS.368L..35L}%
  \BibitemOpen
  \bibfield  {author} {\bibinfo {author} {\bibfnamefont {Y.}~\bibnamefont
  {{Levin}}},\ }\bibfield  {title} {\enquote {\bibinfo {title} {{QPOs during
  magnetar flares are not driven by mechanical normal modes of the crust}},}\
  }\href {\doibase 10.1111/j.1745-3933.2006.00155.x} {\bibfield  {journal}
  {\bibinfo  {journal} {\mnras}\ }\textbf {\bibinfo {volume} {368}},\ \bibinfo
  {pages} {L35--L38} (\bibinfo {year} {2006})},\ \Eprint
  {http://arxiv.org/abs/astro-ph/0601020} {astro-ph/0601020} \BibitemShut
  {NoStop}%
\bibitem [{\citenamefont {{Sotani}}\ \emph
  {et~al.}(2008{\natexlab{a}})\citenamefont {{Sotani}}, \citenamefont
  {{Kokkotas}},\ and\ \citenamefont {{Stergioulas}}}]{2008MNRAS.385L...5S}%
  \BibitemOpen
  \bibfield  {author} {\bibinfo {author} {\bibfnamefont {H.}~\bibnamefont
  {{Sotani}}}, \bibinfo {author} {\bibfnamefont {K.~D.}\ \bibnamefont
  {{Kokkotas}}}, \ and\ \bibinfo {author} {\bibfnamefont {N.}~\bibnamefont
  {{Stergioulas}}},\ }\bibfield  {title} {\enquote {\bibinfo {title}
  {{Alfv{\'e}n quasi-periodic oscillations in magnetars}},}\ }\href {\doibase
  10.1111/j.1745-3933.2007.00420.x} {\bibfield  {journal} {\bibinfo  {journal}
  {\mnras}\ }\textbf {\bibinfo {volume} {385}},\ \bibinfo {pages} {L5--L9}
  (\bibinfo {year} {2008}{\natexlab{a}})},\ \Eprint
  {http://arxiv.org/abs/0710.1113} {arXiv:0710.1113} \BibitemShut {NoStop}%
\bibitem [{\citenamefont {{Gabler}}\ \emph {et~al.}(2011)\citenamefont
  {{Gabler}}, \citenamefont {{Cerd{\'a} Dur{\'a}n}}, \citenamefont {{Font}},
  \citenamefont {{M{\"u}ller}},\ and\ \citenamefont
  {{Stergioulas}}}]{2011MNRAS.410L..37G}%
  \BibitemOpen
  \bibfield  {author} {\bibinfo {author} {\bibfnamefont {M.}~\bibnamefont
  {{Gabler}}}, \bibinfo {author} {\bibfnamefont {P.}~\bibnamefont {{Cerd{\'a}
  Dur{\'a}n}}}, \bibinfo {author} {\bibfnamefont {J.~A.}\ \bibnamefont
  {{Font}}}, \bibinfo {author} {\bibfnamefont {E.}~\bibnamefont
  {{M{\"u}ller}}}, \ and\ \bibinfo {author} {\bibfnamefont {N.}~\bibnamefont
  {{Stergioulas}}},\ }\bibfield  {title} {\enquote {\bibinfo {title}
  {{Magneto-elastic oscillations and the damping of crustal shear modes in
  magnetars}},}\ }\href {\doibase 10.1111/j.1745-3933.2010.00974.x} {\bibfield
  {journal} {\bibinfo  {journal} {\mnras}\ }\textbf {\bibinfo {volume} {410}},\
  \bibinfo {pages} {L37--L41} (\bibinfo {year} {2011})},\ \Eprint
  {http://arxiv.org/abs/1007.0856} {arXiv:1007.0856 [astro-ph.HE]} \BibitemShut
  {NoStop}%
\bibitem [{\citenamefont {{Link}}\ and\ \citenamefont {{van
  Eysden}}(2016)}]{2016ApJ...823L...1L}%
  \BibitemOpen
  \bibfield  {author} {\bibinfo {author} {\bibfnamefont {B.}~\bibnamefont
  {{Link}}}\ and\ \bibinfo {author} {\bibfnamefont {C.~A.}\ \bibnamefont {{van
  Eysden}}},\ }\bibfield  {title} {\enquote {\bibinfo {title} {{Torsional
  Oscillations of a Magnetar with a Tangled Magnetic Field}},}\ }\href
  {\doibase 10.3847/2041-8205/823/1/L1} {\bibfield  {journal} {\bibinfo
  {journal} {\apjl}\ }\textbf {\bibinfo {volume} {823}},\ \bibinfo {eid} {L1}
  (\bibinfo {year} {2016})},\ \Eprint {http://arxiv.org/abs/1604.02372}
  {arXiv:1604.02372 [astro-ph.HE]} \BibitemShut {NoStop}%
\bibitem [{\citenamefont {{Deibel}}\ \emph {et~al.}(2014)\citenamefont
  {{Deibel}}, \citenamefont {{Steiner}},\ and\ \citenamefont
  {{Brown}}}]{2014PhRvC..90b5802D}%
  \BibitemOpen
  \bibfield  {author} {\bibinfo {author} {\bibfnamefont {A.~T.}\ \bibnamefont
  {{Deibel}}}, \bibinfo {author} {\bibfnamefont {A.~W.}\ \bibnamefont
  {{Steiner}}}, \ and\ \bibinfo {author} {\bibfnamefont {E.~F.}\ \bibnamefont
  {{Brown}}},\ }\bibfield  {title} {\enquote {\bibinfo {title} {{Magnetar giant
  flare oscillations and the nuclear symmetry energy}},}\ }\href {\doibase
  10.1103/PhysRevC.90.025802} {\bibfield  {journal} {\bibinfo  {journal}
  {\prc}\ }\textbf {\bibinfo {volume} {90}},\ \bibinfo {eid} {025802} (\bibinfo
  {year} {2014})},\ \Eprint {http://arxiv.org/abs/1303.3270} {arXiv:1303.3270
  [astro-ph.HE]} \BibitemShut {NoStop}%
\bibitem [{\citenamefont {{Gabler}}\ \emph {et~al.}(2018)\citenamefont
  {{Gabler}}, \citenamefont {{Cerd{\'a}-Dur{\'a}n}}, \citenamefont
  {{Stergioulas}}, \citenamefont {{Font}},\ and\ \citenamefont
  {{M{\"u}ller}}}]{2018MNRAS.476.4199G}%
  \BibitemOpen
  \bibfield  {author} {\bibinfo {author} {\bibfnamefont {M.}~\bibnamefont
  {{Gabler}}}, \bibinfo {author} {\bibfnamefont {P.}~\bibnamefont
  {{Cerd{\'a}-Dur{\'a}n}}}, \bibinfo {author} {\bibfnamefont {N.}~\bibnamefont
  {{Stergioulas}}}, \bibinfo {author} {\bibfnamefont {J.~A.}\ \bibnamefont
  {{Font}}}, \ and\ \bibinfo {author} {\bibfnamefont {E.}~\bibnamefont
  {{M{\"u}ller}}},\ }\bibfield  {title} {\enquote {\bibinfo {title}
  {{Constraining properties of high-density matter in neutron stars with
  magneto-elastic oscillations}},}\ }\href {\doibase 10.1093/mnras/sty445}
  {\bibfield  {journal} {\bibinfo  {journal} {\mnras}\ }\textbf {\bibinfo
  {volume} {476}},\ \bibinfo {pages} {4199--4212} (\bibinfo {year} {2018})},\
  \Eprint {http://arxiv.org/abs/1710.02334} {arXiv:1710.02334 [astro-ph.HE]}
  \BibitemShut {NoStop}%
\bibitem [{\citenamefont {{Sotani}}\ \emph {et~al.}(2018)\citenamefont
  {{Sotani}}, \citenamefont {{Iida}},\ and\ \citenamefont
  {{Oyamatsu}}}]{2018MNRAS.479.4735S}%
  \BibitemOpen
  \bibfield  {author} {\bibinfo {author} {\bibfnamefont {H.}~\bibnamefont
  {{Sotani}}}, \bibinfo {author} {\bibfnamefont {K.}~\bibnamefont {{Iida}}}, \
  and\ \bibinfo {author} {\bibfnamefont {K.}~\bibnamefont {{Oyamatsu}}},\
  }\bibfield  {title} {\enquote {\bibinfo {title} {{Constraints on the nuclear
  equation of state and the neutron star structure from crustal torsional
  oscillations}},}\ }\href {\doibase 10.1093/mnras/sty1755} {\bibfield
  {journal} {\bibinfo  {journal} {\mnras}\ }\textbf {\bibinfo {volume} {479}},\
  \bibinfo {pages} {4735--4748} (\bibinfo {year} {2018})},\ \Eprint
  {http://arxiv.org/abs/1807.00528} {arXiv:1807.00528 [astro-ph.HE]}
  \BibitemShut {NoStop}%
\bibitem [{\citenamefont {{Huppenkothen}}\ \emph
  {et~al.}(2014{\natexlab{b}})\citenamefont {{Huppenkothen}}, \citenamefont
  {{Watts}},\ and\ \citenamefont {{Levin}}}]{2014ApJ...793..129H}%
  \BibitemOpen
  \bibfield  {author} {\bibinfo {author} {\bibfnamefont {D.}~\bibnamefont
  {{Huppenkothen}}}, \bibinfo {author} {\bibfnamefont {A.~L.}\ \bibnamefont
  {{Watts}}}, \ and\ \bibinfo {author} {\bibfnamefont {Y.}~\bibnamefont
  {{Levin}}},\ }\bibfield  {title} {\enquote {\bibinfo {title} {{Intermittency
  and Lifetime of the 625 Hz Quasi-periodic Oscillation in the 2004 Hyperflare
  from the Magnetar SGR 1806-20 as Evidence for Magnetic Coupling between the
  Crust and the Core}},}\ }\href {\doibase 10.1088/0004-637X/793/2/129}
  {\bibfield  {journal} {\bibinfo  {journal} {\apj}\ }\textbf {\bibinfo
  {volume} {793}},\ \bibinfo {eid} {129} (\bibinfo {year}
  {2014}{\natexlab{b}})},\ \Eprint {http://arxiv.org/abs/1408.0734}
  {arXiv:1408.0734 [astro-ph.HE]} \BibitemShut {NoStop}%
\bibitem [{\citenamefont {{Pumpe}}\ \emph {et~al.}(2018)\citenamefont
  {{Pumpe}}, \citenamefont {{Gabler}}, \citenamefont {{Steininger}},\ and\
  \citenamefont {{En{\ss}lin}}}]{2018A&A...610A..61P}%
  \BibitemOpen
  \bibfield  {author} {\bibinfo {author} {\bibfnamefont {D.}~\bibnamefont
  {{Pumpe}}}, \bibinfo {author} {\bibfnamefont {M.}~\bibnamefont {{Gabler}}},
  \bibinfo {author} {\bibfnamefont {T.}~\bibnamefont {{Steininger}}}, \ and\
  \bibinfo {author} {\bibfnamefont {T.~A.}\ \bibnamefont {{En{\ss}lin}}},\
  }\bibfield  {title} {\enquote {\bibinfo {title} {{Search for quasi-periodic
  signals in magnetar giant flares. Bayesian inspection of SGR 1806-20 and SGR
  1900+14}},}\ }\href {\doibase 10.1051/0004-6361/201731800} {\bibfield
  {journal} {\bibinfo  {journal} {\aap}\ }\textbf {\bibinfo {volume} {610}},\
  \bibinfo {eid} {A61} (\bibinfo {year} {2018})},\ \Eprint
  {http://arxiv.org/abs/1708.05702} {arXiv:1708.05702 [astro-ph.HE]}
  \BibitemShut {NoStop}%
\bibitem [{\citenamefont {Miller}\ \emph {et~al.}(2018)\citenamefont {Miller},
  \citenamefont {Chirenti},\ and\ \citenamefont {Strohmayer}}]{Miller:2018kmk}%
  \BibitemOpen
  \bibfield  {author} {\bibinfo {author} {\bibfnamefont {M.~Coleman}\
  \bibnamefont {Miller}}, \bibinfo {author} {\bibfnamefont {Cecilia}\
  \bibnamefont {Chirenti}}, \ and\ \bibinfo {author} {\bibfnamefont {Tod}\
  \bibnamefont {Strohmayer}},\ }\bibfield  {title} {\enquote {\bibinfo {title}
  {{A Reanalysis of Quasi-Periodic Oscillations from the SGR 1806-20 Giant
  Flare}},}\ }\href@noop {} {\  (\bibinfo {year} {2018})},\ \Eprint
  {http://arxiv.org/abs/1808.09483} {arXiv:1808.09483 [astro-ph.HE]}
  \BibitemShut {NoStop}%
\bibitem [{\citenamefont {{Sotani}}\ \emph {et~al.}(2007)\citenamefont
  {{Sotani}}, \citenamefont {{Kokkotas}},\ and\ \citenamefont
  {{Stergioulas}}}]{2007MNRAS.375..261S}%
  \BibitemOpen
  \bibfield  {author} {\bibinfo {author} {\bibfnamefont {H.}~\bibnamefont
  {{Sotani}}}, \bibinfo {author} {\bibfnamefont {K.~D.}\ \bibnamefont
  {{Kokkotas}}}, \ and\ \bibinfo {author} {\bibfnamefont {N.}~\bibnamefont
  {{Stergioulas}}},\ }\bibfield  {title} {\enquote {\bibinfo {title}
  {{Torsional oscillations of relativistic stars with dipole magnetic
  fields}},}\ }\href {\doibase 10.1111/j.1365-2966.2006.11304.x} {\bibfield
  {journal} {\bibinfo  {journal} {\mnras}\ }\textbf {\bibinfo {volume} {375}},\
  \bibinfo {pages} {261--277} (\bibinfo {year} {2007})},\ \Eprint
  {http://arxiv.org/abs/astro-ph/0608626} {astro-ph/0608626} \BibitemShut
  {NoStop}%
\bibitem [{\citenamefont {{Tolman}}(1939)}]{1939PhRv...55..364T}%
  \BibitemOpen
  \bibfield  {author} {\bibinfo {author} {\bibfnamefont {R.~C.}\ \bibnamefont
  {{Tolman}}},\ }\bibfield  {title} {\enquote {\bibinfo {title} {{Static
  Solutions of Einstein's Field Equations for Spheres of Fluid}},}\ }\href
  {\doibase 10.1103/PhysRev.55.364} {\bibfield  {journal} {\bibinfo  {journal}
  {Physical Review}\ }\textbf {\bibinfo {volume} {55}},\ \bibinfo {pages}
  {364--373} (\bibinfo {year} {1939})}\BibitemShut {NoStop}%
\bibitem [{\citenamefont {{Oppenheimer}}\ and\ \citenamefont
  {{Volkoff}}(1939)}]{1939PhRv...55..374O}%
  \BibitemOpen
  \bibfield  {author} {\bibinfo {author} {\bibfnamefont {J.~R.}\ \bibnamefont
  {{Oppenheimer}}}\ and\ \bibinfo {author} {\bibfnamefont {G.~M.}\ \bibnamefont
  {{Volkoff}}},\ }\bibfield  {title} {\enquote {\bibinfo {title} {{On Massive
  Neutron Cores}},}\ }\href {\doibase 10.1103/PhysRev.55.374} {\bibfield
  {journal} {\bibinfo  {journal} {Physical Review}\ }\textbf {\bibinfo {volume}
  {55}},\ \bibinfo {pages} {374--381} (\bibinfo {year} {1939})}\BibitemShut
  {NoStop}%
\bibitem [{\citenamefont {{Akmal}}\ \emph {et~al.}(1998)\citenamefont
  {{Akmal}}, \citenamefont {{Pandharipande}},\ and\ \citenamefont
  {{Ravenhall}}}]{1998PhRvC..58.1804A}%
  \BibitemOpen
  \bibfield  {author} {\bibinfo {author} {\bibfnamefont {A.}~\bibnamefont
  {{Akmal}}}, \bibinfo {author} {\bibfnamefont {V.~R.}\ \bibnamefont
  {{Pandharipande}}}, \ and\ \bibinfo {author} {\bibfnamefont {D.~G.}\
  \bibnamefont {{Ravenhall}}},\ }\bibfield  {title} {\enquote {\bibinfo {title}
  {{Equation of state of nucleon matter and neutron star structure}},}\ }\href
  {\doibase 10.1103/PhysRevC.58.1804} {\bibfield  {journal} {\bibinfo
  {journal} {\prc}\ }\textbf {\bibinfo {volume} {58}},\ \bibinfo {pages}
  {1804--1828} (\bibinfo {year} {1998})},\ \Eprint
  {http://arxiv.org/abs/nucl-th/9804027} {nucl-th/9804027} \BibitemShut
  {NoStop}%
\bibitem [{\citenamefont {{Lackey}}\ \emph {et~al.}(2006)\citenamefont
  {{Lackey}}, \citenamefont {{Nayyar}},\ and\ \citenamefont
  {{Owen}}}]{2006PhRvD..73b4021L}%
  \BibitemOpen
  \bibfield  {author} {\bibinfo {author} {\bibfnamefont {B.~D.}\ \bibnamefont
  {{Lackey}}}, \bibinfo {author} {\bibfnamefont {M.}~\bibnamefont {{Nayyar}}},
  \ and\ \bibinfo {author} {\bibfnamefont {B.~J.}\ \bibnamefont {{Owen}}},\
  }\bibfield  {title} {\enquote {\bibinfo {title} {{Observational constraints
  on hyperons in neutron stars}},}\ }\href {\doibase
  10.1103/PhysRevD.73.024021} {\bibfield  {journal} {\bibinfo  {journal}
  {\prd}\ }\textbf {\bibinfo {volume} {73}},\ \bibinfo {eid} {024021} (\bibinfo
  {year} {2006})},\ \Eprint {http://arxiv.org/abs/astro-ph/0507312}
  {astro-ph/0507312} \BibitemShut {NoStop}%
\bibitem [{\citenamefont {{Douchin}}\ and\ \citenamefont
  {{Haensel}}(2001)}]{2001AA...380..151D}%
  \BibitemOpen
  \bibfield  {author} {\bibinfo {author} {\bibfnamefont {F.}~\bibnamefont
  {{Douchin}}}\ and\ \bibinfo {author} {\bibfnamefont {P.}~\bibnamefont
  {{Haensel}}},\ }\bibfield  {title} {\enquote {\bibinfo {title} {{A unified
  equation of state of dense matter and neutron star structure}},}\ }\href
  {\doibase 10.1051/0004-6361:20011402} {\bibfield  {journal} {\bibinfo
  {journal} {\aap}\ }\textbf {\bibinfo {volume} {380}},\ \bibinfo {pages}
  {151--167} (\bibinfo {year} {2001})},\ \Eprint
  {http://arxiv.org/abs/astro-ph/0111092} {astro-ph/0111092} \BibitemShut
  {NoStop}%
\bibitem [{\citenamefont {{Steiner}}(2012)}]{2012PhRvC..85e5804S}%
  \BibitemOpen
  \bibfield  {author} {\bibinfo {author} {\bibfnamefont {A.~W.}\ \bibnamefont
  {{Steiner}}},\ }\bibfield  {title} {\enquote {\bibinfo {title} {{Deep crustal
  heating in a multicomponent accreted neutron star crust}},}\ }\href {\doibase
  10.1103/PhysRevC.85.055804} {\bibfield  {journal} {\bibinfo  {journal}
  {\prc}\ }\textbf {\bibinfo {volume} {85}},\ \bibinfo {eid} {055804} (\bibinfo
  {year} {2012})},\ \Eprint {http://arxiv.org/abs/1202.3378} {arXiv:1202.3378
  [nucl-th]} \BibitemShut {NoStop}%
\bibitem [{\citenamefont {{Negele}}\ and\ \citenamefont
  {{Vautherin}}(1973)}]{1973NuPhA.207..298N}%
  \BibitemOpen
  \bibfield  {author} {\bibinfo {author} {\bibfnamefont {J.~W.}\ \bibnamefont
  {{Negele}}}\ and\ \bibinfo {author} {\bibfnamefont {D.}~\bibnamefont
  {{Vautherin}}},\ }\bibfield  {title} {\enquote {\bibinfo {title} {{Neutron
  star matter at sub-nuclear densities}},}\ }\href {\doibase
  10.1016/0375-9474(73)90349-7} {\bibfield  {journal} {\bibinfo  {journal}
  {Nuclear Physics A}\ }\textbf {\bibinfo {volume} {207}},\ \bibinfo {pages}
  {298--320} (\bibinfo {year} {1973})}\BibitemShut {NoStop}%
\bibitem [{\citenamefont {{Strohmayer}}\ \emph {et~al.}(1991)\citenamefont
  {{Strohmayer}}, \citenamefont {{Ogata}}, \citenamefont {{Iyetomi}},
  \citenamefont {{Ichimaru}},\ and\ \citenamefont {{van
  Horn}}}]{1991ApJ...375..679S}%
  \BibitemOpen
  \bibfield  {author} {\bibinfo {author} {\bibfnamefont {T.}~\bibnamefont
  {{Strohmayer}}}, \bibinfo {author} {\bibfnamefont {S.}~\bibnamefont
  {{Ogata}}}, \bibinfo {author} {\bibfnamefont {H.}~\bibnamefont {{Iyetomi}}},
  \bibinfo {author} {\bibfnamefont {S.}~\bibnamefont {{Ichimaru}}}, \ and\
  \bibinfo {author} {\bibfnamefont {H.~M.}\ \bibnamefont {{van Horn}}},\
  }\bibfield  {title} {\enquote {\bibinfo {title} {{The shear modulus of the
  neutron star crust and nonradial oscillations of neutron stars}},}\ }\href
  {\doibase 10.1086/170231} {\bibfield  {journal} {\bibinfo  {journal} {\apj}\
  }\textbf {\bibinfo {volume} {375}},\ \bibinfo {pages} {679--686} (\bibinfo
  {year} {1991})}\BibitemShut {NoStop}%
\bibitem [{\citenamefont {{Tayler}}(1973)}]{1973MNRAS.161..365T}%
  \BibitemOpen
  \bibfield  {author} {\bibinfo {author} {\bibfnamefont {R.~J.}\ \bibnamefont
  {{Tayler}}},\ }\bibfield  {title} {\enquote {\bibinfo {title} {{The adiabatic
  stability of stars containing magnetic fields-I.Toroidal fields}},}\ }\href
  {\doibase 10.1093/mnras/161.4.365} {\bibfield  {journal} {\bibinfo  {journal}
  {\mnras}\ }\textbf {\bibinfo {volume} {161}},\ \bibinfo {pages} {365}
  (\bibinfo {year} {1973})}\BibitemShut {NoStop}%
\bibitem [{\citenamefont {{Kiuchi}}\ \emph {et~al.}(2008)\citenamefont
  {{Kiuchi}}, \citenamefont {{Shibata}},\ and\ \citenamefont
  {{Yoshida}}}]{2008PhRvD..78b4029K}%
  \BibitemOpen
  \bibfield  {author} {\bibinfo {author} {\bibfnamefont {Kenta}\ \bibnamefont
  {{Kiuchi}}}, \bibinfo {author} {\bibfnamefont {Masaru}\ \bibnamefont
  {{Shibata}}}, \ and\ \bibinfo {author} {\bibfnamefont {Shijun}\ \bibnamefont
  {{Yoshida}}},\ }\bibfield  {title} {\enquote {\bibinfo {title} {{Evolution of
  neutron stars with toroidal magnetic fields: Axisymmetric simulation in full
  general relativity}},}\ }\href {\doibase 10.1103/PhysRevD.78.024029}
  {\bibfield  {journal} {\bibinfo  {journal} {\prd}\ }\textbf {\bibinfo
  {volume} {78}},\ \bibinfo {eid} {024029} (\bibinfo {year} {2008})},\ \Eprint
  {http://arxiv.org/abs/0805.2712} {arXiv:0805.2712 [astro-ph]} \BibitemShut
  {NoStop}%
\bibitem [{\citenamefont {{Kiuchi}}\ \emph {et~al.}(2011)\citenamefont
  {{Kiuchi}}, \citenamefont {{Yoshida}},\ and\ \citenamefont
  {{Shibata}}}]{2011A&A...532A..30K}%
  \BibitemOpen
  \bibfield  {author} {\bibinfo {author} {\bibfnamefont {K.}~\bibnamefont
  {{Kiuchi}}}, \bibinfo {author} {\bibfnamefont {S.}~\bibnamefont {{Yoshida}}},
  \ and\ \bibinfo {author} {\bibfnamefont {M.}~\bibnamefont {{Shibata}}},\
  }\bibfield  {title} {\enquote {\bibinfo {title} {{Nonaxisymmetric
  instabilities of neutron star with toroidal magnetic fields}},}\ }\href
  {\doibase 10.1051/0004-6361/201016242} {\bibfield  {journal} {\bibinfo
  {journal} {\aap}\ }\textbf {\bibinfo {volume} {532}},\ \bibinfo {eid} {A30}
  (\bibinfo {year} {2011})},\ \Eprint {http://arxiv.org/abs/1104.5561}
  {arXiv:1104.5561 [astro-ph.HE]} \BibitemShut {NoStop}%
\bibitem [{\citenamefont {{Markey}}\ and\ \citenamefont
  {{Tayler}}(1973)}]{1973MNRAS.163...77M}%
  \BibitemOpen
  \bibfield  {author} {\bibinfo {author} {\bibfnamefont {P.}~\bibnamefont
  {{Markey}}}\ and\ \bibinfo {author} {\bibfnamefont {R.~J.}\ \bibnamefont
  {{Tayler}}},\ }\bibfield  {title} {\enquote {\bibinfo {title} {{The adiabatic
  stability of stars containing magnetic fields. II. Poloidal fields}},}\
  }\href {\doibase 10.1093/mnras/163.1.77} {\bibfield  {journal} {\bibinfo
  {journal} {\mnras}\ }\textbf {\bibinfo {volume} {163}},\ \bibinfo {pages}
  {77--91} (\bibinfo {year} {1973})}\BibitemShut {NoStop}%
\bibitem [{\citenamefont {{Markey}}\ and\ \citenamefont
  {{Tayler}}(1974)}]{1974MNRAS.168..505M}%
  \BibitemOpen
  \bibfield  {author} {\bibinfo {author} {\bibfnamefont {P.}~\bibnamefont
  {{Markey}}}\ and\ \bibinfo {author} {\bibfnamefont {R.~J.}\ \bibnamefont
  {{Tayler}}},\ }\bibfield  {title} {\enquote {\bibinfo {title} {{The adiabatic
  stability of stars containing magnetic fields-III. Additional results for
  poloidal fields}},}\ }\href {\doibase 10.1093/mnras/168.3.505} {\bibfield
  {journal} {\bibinfo  {journal} {\mnras}\ }\textbf {\bibinfo {volume} {168}},\
  \bibinfo {pages} {505--514} (\bibinfo {year} {1974})}\BibitemShut {NoStop}%
\bibitem [{\citenamefont {{Flowers}}\ and\ \citenamefont
  {{Ruderman}}(1977)}]{1977ApJ...215..302F}%
  \BibitemOpen
  \bibfield  {author} {\bibinfo {author} {\bibfnamefont {E.}~\bibnamefont
  {{Flowers}}}\ and\ \bibinfo {author} {\bibfnamefont {M.~A.}\ \bibnamefont
  {{Ruderman}}},\ }\bibfield  {title} {\enquote {\bibinfo {title} {{Evolution
  of pulsar magnetic fields}},}\ }\href {\doibase 10.1086/155359} {\bibfield
  {journal} {\bibinfo  {journal} {\apj}\ }\textbf {\bibinfo {volume} {215}},\
  \bibinfo {pages} {302--310} (\bibinfo {year} {1977})}\BibitemShut {NoStop}%
\bibitem [{\citenamefont {{Lasky}}\ \emph {et~al.}(2011)\citenamefont
  {{Lasky}}, \citenamefont {{Zink}}, \citenamefont {{Kokkotas}},\ and\
  \citenamefont {{Glampedakis}}}]{2011ApJ...735L..20L}%
  \BibitemOpen
  \bibfield  {author} {\bibinfo {author} {\bibfnamefont {Paul~D.}\ \bibnamefont
  {{Lasky}}}, \bibinfo {author} {\bibfnamefont {Burkhard}\ \bibnamefont
  {{Zink}}}, \bibinfo {author} {\bibfnamefont {Kostas~D.}\ \bibnamefont
  {{Kokkotas}}}, \ and\ \bibinfo {author} {\bibfnamefont {Kostas}\ \bibnamefont
  {{Glampedakis}}},\ }\bibfield  {title} {\enquote {\bibinfo {title}
  {{Hydromagnetic Instabilities in Relativistic Neutron Stars}},}\ }\href
  {\doibase 10.1088/2041-8205/735/1/L20} {\bibfield  {journal} {\bibinfo
  {journal} {\apjl}\ }\textbf {\bibinfo {volume} {735}},\ \bibinfo {eid} {L20}
  (\bibinfo {year} {2011})},\ \Eprint {http://arxiv.org/abs/1105.1895}
  {arXiv:1105.1895 [astro-ph.SR]} \BibitemShut {NoStop}%
\bibitem [{\citenamefont {{Ciolfi}}\ and\ \citenamefont
  {{Rezzolla}}(2012)}]{2012ApJ...760....1C}%
  \BibitemOpen
  \bibfield  {author} {\bibinfo {author} {\bibfnamefont {Riccardo}\
  \bibnamefont {{Ciolfi}}}\ and\ \bibinfo {author} {\bibfnamefont {Luciano}\
  \bibnamefont {{Rezzolla}}},\ }\bibfield  {title} {\enquote {\bibinfo {title}
  {{Poloidal-field Instability in Magnetized Relativistic Stars}},}\ }\href
  {\doibase 10.1088/0004-637X/760/1/1} {\bibfield  {journal} {\bibinfo
  {journal} {\apj}\ }\textbf {\bibinfo {volume} {760}},\ \bibinfo {eid} {1}
  (\bibinfo {year} {2012})},\ \Eprint {http://arxiv.org/abs/1206.6604}
  {arXiv:1206.6604 [astro-ph.SR]} \BibitemShut {NoStop}%
\bibitem [{\citenamefont {{Braithwaite}}\ and\ \citenamefont
  {{Nordlund}}(2006)}]{2006A&A...450.1077B}%
  \BibitemOpen
  \bibfield  {author} {\bibinfo {author} {\bibfnamefont {J.}~\bibnamefont
  {{Braithwaite}}}\ and\ \bibinfo {author} {\bibfnamefont {{\AA}.}~\bibnamefont
  {{Nordlund}}},\ }\bibfield  {title} {\enquote {\bibinfo {title} {{Stable
  magnetic fields in stellar interiors}},}\ }\href {\doibase
  10.1051/0004-6361:20041980} {\bibfield  {journal} {\bibinfo  {journal}
  {\aap}\ }\textbf {\bibinfo {volume} {450}},\ \bibinfo {pages} {1077--1095}
  (\bibinfo {year} {2006})},\ \Eprint {http://arxiv.org/abs/astro-ph/0510316}
  {astro-ph/0510316} \BibitemShut {NoStop}%
\bibitem [{\citenamefont {{Braithwaite}}(2008)}]{2008MNRAS.386.1947B}%
  \BibitemOpen
  \bibfield  {author} {\bibinfo {author} {\bibfnamefont {J.}~\bibnamefont
  {{Braithwaite}}},\ }\bibfield  {title} {\enquote {\bibinfo {title} {{On
  non-axisymmetric magnetic equilibria in stars}},}\ }\href {\doibase
  10.1111/j.1365-2966.2008.13218.x} {\bibfield  {journal} {\bibinfo  {journal}
  {\mnras}\ }\textbf {\bibinfo {volume} {386}},\ \bibinfo {pages} {1947--1958}
  (\bibinfo {year} {2008})},\ \Eprint {http://arxiv.org/abs/0803.1661}
  {arXiv:0803.1661} \BibitemShut {NoStop}%
\bibitem [{\citenamefont {Ioka}\ and\ \citenamefont
  {Sasaki}(2004)}]{Ioka:2003nh}%
  \BibitemOpen
  \bibfield  {author} {\bibinfo {author} {\bibfnamefont {Kunihito}\
  \bibnamefont {Ioka}}\ and\ \bibinfo {author} {\bibfnamefont {Misao}\
  \bibnamefont {Sasaki}},\ }\bibfield  {title} {\enquote {\bibinfo {title}
  {{Relativistic stars with poloidal and toroidal magnetic fields and
  meridional flow}},}\ }\href {\doibase 10.1086/379650} {\bibfield  {journal}
  {\bibinfo  {journal} {Astrophys. J.}\ }\textbf {\bibinfo {volume} {600}},\
  \bibinfo {pages} {296--316} (\bibinfo {year} {2004})},\ \Eprint
  {http://arxiv.org/abs/astro-ph/0305352} {arXiv:astro-ph/0305352 [astro-ph]}
  \BibitemShut {NoStop}%
\bibitem [{\citenamefont {{Colaiuda}}\ \emph {et~al.}(2008)\citenamefont
  {{Colaiuda}}, \citenamefont {{Ferrari}}, \citenamefont {{Gualtieri}},\ and\
  \citenamefont {{Pons}}}]{2008MNRAS.385.2080C}%
  \BibitemOpen
  \bibfield  {author} {\bibinfo {author} {\bibfnamefont {A.}~\bibnamefont
  {{Colaiuda}}}, \bibinfo {author} {\bibfnamefont {V.}~\bibnamefont
  {{Ferrari}}}, \bibinfo {author} {\bibfnamefont {L.}~\bibnamefont
  {{Gualtieri}}}, \ and\ \bibinfo {author} {\bibfnamefont {J.~A.}\ \bibnamefont
  {{Pons}}},\ }\bibfield  {title} {\enquote {\bibinfo {title} {{Relativistic
  models of magnetars: structure and deformations}},}\ }\href {\doibase
  10.1111/j.1365-2966.2008.12966.x} {\bibfield  {journal} {\bibinfo  {journal}
  {\mnras}\ }\textbf {\bibinfo {volume} {385}},\ \bibinfo {pages} {2080--2096}
  (\bibinfo {year} {2008})},\ \Eprint {http://arxiv.org/abs/0712.2162}
  {arXiv:0712.2162} \BibitemShut {NoStop}%
\bibitem [{\citenamefont {{Chirenti}}\ and\ \citenamefont
  {{Sk{\'a}kala}}(2013)}]{2013PhRvD..88j4018C}%
  \BibitemOpen
  \bibfield  {author} {\bibinfo {author} {\bibfnamefont {C.}~\bibnamefont
  {{Chirenti}}}\ and\ \bibinfo {author} {\bibfnamefont {J.}~\bibnamefont
  {{Sk{\'a}kala}}},\ }\bibfield  {title} {\enquote {\bibinfo {title} {{Effect
  of magnetic fields on the r-modes of slowly rotating relativistic neutron
  stars}},}\ }\href {\doibase 10.1103/PhysRevD.88.104018} {\bibfield  {journal}
  {\bibinfo  {journal} {\prd}\ }\textbf {\bibinfo {volume} {88}},\ \bibinfo
  {eid} {104018} (\bibinfo {year} {2013})},\ \Eprint
  {http://arxiv.org/abs/1308.3685} {arXiv:1308.3685 [gr-qc]} \BibitemShut
  {NoStop}%
\bibitem [{\citenamefont {{Pili}}\ \emph {et~al.}(2015)\citenamefont {{Pili}},
  \citenamefont {{Bucciantini}},\ and\ \citenamefont {{Del
  Zanna}}}]{2015MNRAS.447.2821P}%
  \BibitemOpen
  \bibfield  {author} {\bibinfo {author} {\bibfnamefont {A.~G.}\ \bibnamefont
  {{Pili}}}, \bibinfo {author} {\bibfnamefont {N.}~\bibnamefont
  {{Bucciantini}}}, \ and\ \bibinfo {author} {\bibfnamefont {L.}~\bibnamefont
  {{Del Zanna}}},\ }\bibfield  {title} {\enquote {\bibinfo {title} {{General
  relativistic neutron stars with twisted magnetosphere}},}\ }\href {\doibase
  10.1093/mnras/stu2628} {\bibfield  {journal} {\bibinfo  {journal} {\mnras}\
  }\textbf {\bibinfo {volume} {447}},\ \bibinfo {pages} {2821--2835} (\bibinfo
  {year} {2015})},\ \Eprint {http://arxiv.org/abs/1412.4036} {arXiv:1412.4036
  [astro-ph.HE]} \BibitemShut {NoStop}%
\bibitem [{\citenamefont {{Akg{\"u}n}}\ \emph {et~al.}(2018)\citenamefont
  {{Akg{\"u}n}}, \citenamefont {{Cerd{\'a}-Dur{\'a}n}}, \citenamefont
  {{Miralles}},\ and\ \citenamefont {{Pons}}}]{2018MNRAS.474..625A}%
  \BibitemOpen
  \bibfield  {author} {\bibinfo {author} {\bibfnamefont {T.}~\bibnamefont
  {{Akg{\"u}n}}}, \bibinfo {author} {\bibfnamefont {P.}~\bibnamefont
  {{Cerd{\'a}-Dur{\'a}n}}}, \bibinfo {author} {\bibfnamefont {J.~A.}\
  \bibnamefont {{Miralles}}}, \ and\ \bibinfo {author} {\bibfnamefont {J.~A.}\
  \bibnamefont {{Pons}}},\ }\bibfield  {title} {\enquote {\bibinfo {title}
  {{The force-free twisted magnetosphere of a neutron star - II. Degeneracies
  of the Grad-Shafranov equation}},}\ }\href {\doibase 10.1093/mnras/stx2814}
  {\bibfield  {journal} {\bibinfo  {journal} {\mnras}\ }\textbf {\bibinfo
  {volume} {474}},\ \bibinfo {pages} {625--635} (\bibinfo {year} {2018})},\
  \Eprint {http://arxiv.org/abs/1709.10044} {arXiv:1709.10044 [astro-ph.HE]}
  \BibitemShut {NoStop}%
\bibitem [{\citenamefont {{Ciolfi}}\ \emph {et~al.}(2009)\citenamefont
  {{Ciolfi}}, \citenamefont {{Ferrari}}, \citenamefont {{Gualtieri}},\ and\
  \citenamefont {{Pons}}}]{2009MNRAS.397..913C}%
  \BibitemOpen
  \bibfield  {author} {\bibinfo {author} {\bibfnamefont {R.}~\bibnamefont
  {{Ciolfi}}}, \bibinfo {author} {\bibfnamefont {V.}~\bibnamefont {{Ferrari}}},
  \bibinfo {author} {\bibfnamefont {L.}~\bibnamefont {{Gualtieri}}}, \ and\
  \bibinfo {author} {\bibfnamefont {J.~A.}\ \bibnamefont {{Pons}}},\ }\bibfield
   {title} {\enquote {\bibinfo {title} {{Relativistic models of magnetars: the
  twisted torus magnetic field configuration}},}\ }\href {\doibase
  10.1111/j.1365-2966.2009.14990.x} {\bibfield  {journal} {\bibinfo  {journal}
  {\mnras}\ }\textbf {\bibinfo {volume} {397}},\ \bibinfo {pages} {913--924}
  (\bibinfo {year} {2009})},\ \Eprint {http://arxiv.org/abs/0903.0556}
  {arXiv:0903.0556 [astro-ph.SR]} \BibitemShut {NoStop}%
\bibitem [{\citenamefont {{Cowling}}(1941)}]{1941MNRAS.101..367C}%
  \BibitemOpen
  \bibfield  {author} {\bibinfo {author} {\bibfnamefont {T.~G.}\ \bibnamefont
  {{Cowling}}},\ }\bibfield  {title} {\enquote {\bibinfo {title} {{The
  non-radial oscillations of polytropic stars}},}\ }\href {\doibase
  10.1093/mnras/101.8.367} {\bibfield  {journal} {\bibinfo  {journal} {\mnras}\
  }\textbf {\bibinfo {volume} {101}},\ \bibinfo {pages} {367} (\bibinfo {year}
  {1941})}\BibitemShut {NoStop}%
\bibitem [{\citenamefont {{Kokkotas}}\ and\ \citenamefont
  {{Schmidt}}(1999)}]{1999LRR.....2....2K}%
  \BibitemOpen
  \bibfield  {author} {\bibinfo {author} {\bibfnamefont {K.~D.}\ \bibnamefont
  {{Kokkotas}}}\ and\ \bibinfo {author} {\bibfnamefont {B.~G.}\ \bibnamefont
  {{Schmidt}}},\ }\bibfield  {title} {\enquote {\bibinfo {title} {{Quasi-Normal
  Modes of Stars and Black Holes}},}\ }\href {\doibase 10.12942/lrr-1999-2}
  {\bibfield  {journal} {\bibinfo  {journal} {Living Reviews in Relativity}\
  }\textbf {\bibinfo {volume} {2}},\ \bibinfo {eid} {2} (\bibinfo {year}
  {1999})},\ \Eprint {http://arxiv.org/abs/gr-qc/9909058} {gr-qc/9909058}
  \BibitemShut {NoStop}%
\bibitem [{\citenamefont {{Messios}}\ \emph {et~al.}(2001)\citenamefont
  {{Messios}}, \citenamefont {{Papadopoulos}},\ and\ \citenamefont
  {{Stergioulas}}}]{2001MNRAS.328.1161M}%
  \BibitemOpen
  \bibfield  {author} {\bibinfo {author} {\bibfnamefont {N.}~\bibnamefont
  {{Messios}}}, \bibinfo {author} {\bibfnamefont {D.~B.}\ \bibnamefont
  {{Papadopoulos}}}, \ and\ \bibinfo {author} {\bibfnamefont {N.}~\bibnamefont
  {{Stergioulas}}},\ }\bibfield  {title} {\enquote {\bibinfo {title}
  {{Torsional oscillations of magnetized relativistic stars}},}\ }\href
  {\doibase 10.1046/j.1365-8711.2001.04645.x} {\bibfield  {journal} {\bibinfo
  {journal} {\mnras}\ }\textbf {\bibinfo {volume} {328}},\ \bibinfo {pages}
  {1161--1168} (\bibinfo {year} {2001})},\ \Eprint
  {http://arxiv.org/abs/astro-ph/0105175} {astro-ph/0105175} \BibitemShut
  {NoStop}%
\bibitem [{\citenamefont {{Colaiuda}}\ and\ \citenamefont
  {{Kokkotas}}(2012)}]{2012MNRAS.423..811C}%
  \BibitemOpen
  \bibfield  {author} {\bibinfo {author} {\bibfnamefont {A.}~\bibnamefont
  {{Colaiuda}}}\ and\ \bibinfo {author} {\bibfnamefont {K.~D.}\ \bibnamefont
  {{Kokkotas}}},\ }\bibfield  {title} {\enquote {\bibinfo {title} {{Coupled
  polar-axial magnetar oscillations}},}\ }\href {\doibase
  10.1111/j.1365-2966.2012.20919.x} {\bibfield  {journal} {\bibinfo  {journal}
  {\mnras}\ }\textbf {\bibinfo {volume} {423}},\ \bibinfo {pages} {811--821}
  (\bibinfo {year} {2012})},\ \Eprint {http://arxiv.org/abs/1112.3561}
  {arXiv:1112.3561 [astro-ph.HE]} \BibitemShut {NoStop}%
\bibitem [{\citenamefont {{Colaiuda}}\ \emph {et~al.}(2009)\citenamefont
  {{Colaiuda}}, \citenamefont {{Beyer}},\ and\ \citenamefont
  {{Kokkotas}}}]{2009MNRAS.396.1441C}%
  \BibitemOpen
  \bibfield  {author} {\bibinfo {author} {\bibfnamefont {A.}~\bibnamefont
  {{Colaiuda}}}, \bibinfo {author} {\bibfnamefont {H.}~\bibnamefont {{Beyer}}},
  \ and\ \bibinfo {author} {\bibfnamefont {K.~D.}\ \bibnamefont {{Kokkotas}}},\
  }\bibfield  {title} {\enquote {\bibinfo {title} {{On the quasi-periodic
  oscillations in magnetars}},}\ }\href {\doibase
  10.1111/j.1365-2966.2009.14878.x} {\bibfield  {journal} {\bibinfo  {journal}
  {\mnras}\ }\textbf {\bibinfo {volume} {396}},\ \bibinfo {pages} {1441--1448}
  (\bibinfo {year} {2009})},\ \Eprint {http://arxiv.org/abs/0902.1401}
  {arXiv:0902.1401 [astro-ph.HE]} \BibitemShut {NoStop}%
\bibitem [{\citenamefont {{van Hoven}}\ and\ \citenamefont
  {{Levin}}(2011)}]{2011MNRAS.410.1036V}%
  \BibitemOpen
  \bibfield  {author} {\bibinfo {author} {\bibfnamefont {M.}~\bibnamefont {{van
  Hoven}}}\ and\ \bibinfo {author} {\bibfnamefont {Y.}~\bibnamefont
  {{Levin}}},\ }\bibfield  {title} {\enquote {\bibinfo {title} {{Magnetar
  oscillations - I. Strongly coupled dynamics of the crust and the core}},}\
  }\href {\doibase 10.1111/j.1365-2966.2010.17499.x} {\bibfield  {journal}
  {\bibinfo  {journal} {\mnras}\ }\textbf {\bibinfo {volume} {410}},\ \bibinfo
  {pages} {1036--1051} (\bibinfo {year} {2011})},\ \Eprint
  {http://arxiv.org/abs/1006.0348} {arXiv:1006.0348 [astro-ph.HE]} \BibitemShut
  {NoStop}%
\bibitem [{\citenamefont {{Colaiuda}}\ and\ \citenamefont
  {{Kokkotas}}(2011)}]{2011MNRAS.414.3014C}%
  \BibitemOpen
  \bibfield  {author} {\bibinfo {author} {\bibfnamefont {A.}~\bibnamefont
  {{Colaiuda}}}\ and\ \bibinfo {author} {\bibfnamefont {K.~D.}\ \bibnamefont
  {{Kokkotas}}},\ }\bibfield  {title} {\enquote {\bibinfo {title} {{Magnetar
  oscillations in the presence of a crust}},}\ }\href {\doibase
  10.1111/j.1365-2966.2011.18602.x} {\bibfield  {journal} {\bibinfo  {journal}
  {\mnras}\ }\textbf {\bibinfo {volume} {414}},\ \bibinfo {pages} {3014--3022}
  (\bibinfo {year} {2011})},\ \Eprint {http://arxiv.org/abs/1012.3103}
  {arXiv:1012.3103 [gr-qc]} \BibitemShut {NoStop}%
\bibitem [{\citenamefont {{Sotani}}\ \emph
  {et~al.}(2008{\natexlab{b}})\citenamefont {{Sotani}}, \citenamefont
  {{Colaiuda}},\ and\ \citenamefont {{Kokkotas}}}]{2008MNRAS.385.2161S}%
  \BibitemOpen
  \bibfield  {author} {\bibinfo {author} {\bibfnamefont {H.}~\bibnamefont
  {{Sotani}}}, \bibinfo {author} {\bibfnamefont {A.}~\bibnamefont
  {{Colaiuda}}}, \ and\ \bibinfo {author} {\bibfnamefont {K.~D.}\ \bibnamefont
  {{Kokkotas}}},\ }\bibfield  {title} {\enquote {\bibinfo {title} {{Constraints
  on the magnetic field geometry of magnetars}},}\ }\href {\doibase
  10.1111/j.1365-2966.2008.12977.x} {\bibfield  {journal} {\bibinfo  {journal}
  {\mnras}\ }\textbf {\bibinfo {volume} {385}},\ \bibinfo {pages} {2161--2165}
  (\bibinfo {year} {2008}{\natexlab{b}})},\ \Eprint
  {http://arxiv.org/abs/0711.1518} {arXiv:0711.1518 [gr-qc]} \BibitemShut
  {NoStop}%
\bibitem [{\citenamefont {{Schumaker}}\ and\ \citenamefont
  {{Thorne}}(1983)}]{1983MNRAS.203..457S}%
  \BibitemOpen
  \bibfield  {author} {\bibinfo {author} {\bibfnamefont {B.~L.}\ \bibnamefont
  {{Schumaker}}}\ and\ \bibinfo {author} {\bibfnamefont {K.~S.}\ \bibnamefont
  {{Thorne}}},\ }\bibfield  {title} {\enquote {\bibinfo {title} {{Torsional
  oscillations of neutron stars}},}\ }\href {\doibase 10.1093/mnras/203.2.457}
  {\bibfield  {journal} {\bibinfo  {journal} {\mnras}\ }\textbf {\bibinfo
  {volume} {203}},\ \bibinfo {pages} {457--489} (\bibinfo {year}
  {1983})}\BibitemShut {NoStop}%
\bibitem [{\citenamefont {{Gabler}}\ \emph {et~al.}(2013)\citenamefont
  {{Gabler}}, \citenamefont {{Cerd{\'a}-Dur{\'a}n}}, \citenamefont {{Font}},
  \citenamefont {{M{\"u}ller}},\ and\ \citenamefont
  {{Stergioulas}}}]{2013MNRAS.430.1811G}%
  \BibitemOpen
  \bibfield  {author} {\bibinfo {author} {\bibfnamefont {M.}~\bibnamefont
  {{Gabler}}}, \bibinfo {author} {\bibfnamefont {P.}~\bibnamefont
  {{Cerd{\'a}-Dur{\'a}n}}}, \bibinfo {author} {\bibfnamefont {J.~A.}\
  \bibnamefont {{Font}}}, \bibinfo {author} {\bibfnamefont {E.}~\bibnamefont
  {{M{\"u}ller}}}, \ and\ \bibinfo {author} {\bibfnamefont {N.}~\bibnamefont
  {{Stergioulas}}},\ }\bibfield  {title} {\enquote {\bibinfo {title}
  {{Magneto-elastic oscillations of neutron stars: exploring different magnetic
  field configurations}},}\ }\href {\doibase 10.1093/mnras/sts721} {\bibfield
  {journal} {\bibinfo  {journal} {\mnras}\ }\textbf {\bibinfo {volume} {430}},\
  \bibinfo {pages} {1811--1831} (\bibinfo {year} {2013})},\ \Eprint
  {http://arxiv.org/abs/1208.6443} {arXiv:1208.6443 [astro-ph.SR]} \BibitemShut
  {NoStop}%
\bibitem [{\citenamefont {{Chirenti}}\ \emph {et~al.}(2015)\citenamefont
  {{Chirenti}}, \citenamefont {{de Souza}},\ and\ \citenamefont
  {{Kastaun}}}]{2015PhRvD..91d4034C}%
  \BibitemOpen
  \bibfield  {author} {\bibinfo {author} {\bibfnamefont {C.}~\bibnamefont
  {{Chirenti}}}, \bibinfo {author} {\bibfnamefont {G.~H.}\ \bibnamefont {{de
  Souza}}}, \ and\ \bibinfo {author} {\bibfnamefont {W.}~\bibnamefont
  {{Kastaun}}},\ }\bibfield  {title} {\enquote {\bibinfo {title} {{Fundamental
  oscillation modes of neutron stars: Validity of universal relations}},}\
  }\href {\doibase 10.1103/PhysRevD.91.044034} {\bibfield  {journal} {\bibinfo
  {journal} {\prd}\ }\textbf {\bibinfo {volume} {91}},\ \bibinfo {eid} {044034}
  (\bibinfo {year} {2015})},\ \Eprint {http://arxiv.org/abs/1501.02970}
  {arXiv:1501.02970 [gr-qc]} \BibitemShut {NoStop}%
\bibitem [{\citenamefont {{Yagi}}\ and\ \citenamefont
  {{Yunes}}(2013)}]{2013PhRvD..88b3009Y}%
  \BibitemOpen
  \bibfield  {author} {\bibinfo {author} {\bibfnamefont {K.}~\bibnamefont
  {{Yagi}}}\ and\ \bibinfo {author} {\bibfnamefont {N.}~\bibnamefont
  {{Yunes}}},\ }\bibfield  {title} {\enquote {\bibinfo {title} {{I-Love-Q
  relations in neutron stars and their applications to astrophysics,
  gravitational waves, and fundamental physics}},}\ }\href {\doibase
  10.1103/PhysRevD.88.023009} {\bibfield  {journal} {\bibinfo  {journal}
  {\prd}\ }\textbf {\bibinfo {volume} {88}},\ \bibinfo {eid} {023009} (\bibinfo
  {year} {2013})},\ \Eprint {http://arxiv.org/abs/1303.1528} {arXiv:1303.1528
  [gr-qc]} \BibitemShut {NoStop}%
\bibitem [{\citenamefont {{Corsi}}\ and\ \citenamefont
  {{Owen}}(2011)}]{2011PhRvD..83j4014C}%
  \BibitemOpen
  \bibfield  {author} {\bibinfo {author} {\bibfnamefont {A.}~\bibnamefont
  {{Corsi}}}\ and\ \bibinfo {author} {\bibfnamefont {B.~J.}\ \bibnamefont
  {{Owen}}},\ }\bibfield  {title} {\enquote {\bibinfo {title} {{Maximum
  gravitational-wave energy emissible in magnetar flares}},}\ }\href {\doibase
  10.1103/PhysRevD.83.104014} {\bibfield  {journal} {\bibinfo  {journal}
  {\prd}\ }\textbf {\bibinfo {volume} {83}},\ \bibinfo {eid} {104014} (\bibinfo
  {year} {2011})},\ \Eprint {http://arxiv.org/abs/1102.3421} {arXiv:1102.3421
  [gr-qc]} \BibitemShut {NoStop}%
\bibitem [{\citenamefont {{Quitzow-James}}\ \emph {et~al.}(2017)\citenamefont
  {{Quitzow-James}}, \citenamefont {{Brau}}, \citenamefont {{Clark}},
  \citenamefont {{Coughlin}}, \citenamefont {{Coughlin}}, \citenamefont
  {{Frey}}, \citenamefont {{Schale}}, \citenamefont {{Talukder}},\ and\
  \citenamefont {{Thrane}}}]{2017CQGra..34p4002Q}%
  \BibitemOpen
  \bibfield  {author} {\bibinfo {author} {\bibfnamefont {R.}~\bibnamefont
  {{Quitzow-James}}}, \bibinfo {author} {\bibfnamefont {J.}~\bibnamefont
  {{Brau}}}, \bibinfo {author} {\bibfnamefont {J.~A.}\ \bibnamefont {{Clark}}},
  \bibinfo {author} {\bibfnamefont {M.~W.}\ \bibnamefont {{Coughlin}}},
  \bibinfo {author} {\bibfnamefont {S.~B.}\ \bibnamefont {{Coughlin}}},
  \bibinfo {author} {\bibfnamefont {R.}~\bibnamefont {{Frey}}}, \bibinfo
  {author} {\bibfnamefont {P.}~\bibnamefont {{Schale}}}, \bibinfo {author}
  {\bibfnamefont {D.}~\bibnamefont {{Talukder}}}, \ and\ \bibinfo {author}
  {\bibfnamefont {E.}~\bibnamefont {{Thrane}}},\ }\bibfield  {title} {\enquote
  {\bibinfo {title} {{Exploring a search for long-duration transient
  gravitational waves associated with magnetar bursts}},}\ }\href {\doibase
  10.1088/1361-6382/aa7d5b} {\bibfield  {journal} {\bibinfo  {journal}
  {Classical and Quantum Gravity}\ }\textbf {\bibinfo {volume} {34}},\ \bibinfo
  {eid} {164002} (\bibinfo {year} {2017})},\ \Eprint
  {http://arxiv.org/abs/1704.03979} {arXiv:1704.03979 [astro-ph.IM]}
  \BibitemShut {NoStop}%
\bibitem [{\citenamefont {{Levin}}\ and\ \citenamefont {{van
  Hoven}}(2011)}]{2011MNRAS.418..659L}%
  \BibitemOpen
  \bibfield  {author} {\bibinfo {author} {\bibfnamefont {Y.}~\bibnamefont
  {{Levin}}}\ and\ \bibinfo {author} {\bibfnamefont {M.}~\bibnamefont {{van
  Hoven}}},\ }\bibfield  {title} {\enquote {\bibinfo {title} {{On the
  excitation of f modes and torsional modes by magnetar giant flares}},}\
  }\href {\doibase 10.1111/j.1365-2966.2011.19515.x} {\bibfield  {journal}
  {\bibinfo  {journal} {\mnras}\ }\textbf {\bibinfo {volume} {418}},\ \bibinfo
  {pages} {659--663} (\bibinfo {year} {2011})},\ \Eprint
  {http://arxiv.org/abs/1103.0880} {arXiv:1103.0880 [astro-ph.HE]} \BibitemShut
  {NoStop}%
\bibitem [{\citenamefont {{Punturo}}\ \emph {et~al.}(2010)\citenamefont
  {{Punturo}} \emph {et~al.}}]{2010CQGra..27s4002P}%
  \BibitemOpen
  \bibfield  {author} {\bibinfo {author} {\bibfnamefont {M.}~\bibnamefont
  {{Punturo}}} \emph {et~al.},\ }\bibfield  {title} {\enquote {\bibinfo {title}
  {{The Einstein Telescope: a third-generation gravitational wave
  observatory}},}\ }\href {\doibase 10.1088/0264-9381/27/19/194002} {\bibfield
  {journal} {\bibinfo  {journal} {Classical and Quantum Gravity}\ }\textbf
  {\bibinfo {volume} {27}},\ \bibinfo {eid} {194002} (\bibinfo {year}
  {2010})}\BibitemShut {NoStop}%
\bibitem [{\citenamefont {{Abbott}}\ \emph {et~al.}(2017)\citenamefont
  {{Abbott}} \emph {et~al.}}]{2017CQGra..34d4001A}%
  \BibitemOpen
  \bibfield  {author} {\bibinfo {author} {\bibfnamefont {B.~P.}\ \bibnamefont
  {{Abbott}}} \emph {et~al.},\ }\bibfield  {title} {\enquote {\bibinfo {title}
  {{Exploring the sensitivity of next generation gravitational wave
  detectors}},}\ }\href {\doibase 10.1088/1361-6382/aa51f4} {\bibfield
  {journal} {\bibinfo  {journal} {Classical and Quantum Gravity}\ }\textbf
  {\bibinfo {volume} {34}},\ \bibinfo {eid} {044001} (\bibinfo {year}
  {2017})},\ \Eprint {http://arxiv.org/abs/1607.08697} {arXiv:1607.08697
  [astro-ph.IM]} \BibitemShut {NoStop}%
\end{thebibliography}%
\end{document}